\documentclass[11pt,a4paper]{article}
\usepackage{jheppub}
\usepackage[T1]{fontenc}
\usepackage{xcolor}
\usepackage{bm}		    
\usepackage{pdfcolmk}
\usepackage{amsmath}
\usepackage{amssymb}
\usepackage{graphicx}
\usepackage{esint}
\usepackage{array}
\usepackage{cancel}

\PassOptionsToPackage{normalem}{ulem}
\usepackage{ulem}

\makeatletter

\newcommand{\eref}[1]{eq.\,(\ref{#1})}
\newcommand{\erefs}[2]{eqs.\,(\ref{#1}) and (\ref{#2})}
\newcommand{\erefss}[3]{eqs.\,(\ref{#1}), (\ref{#2}), and (\ref{#3})}
\newcommand{\erefsss}[4]{eqs.\,(\ref{#1}), (\ref{#2}), (\ref{#3}), and (\ref{#4})}
\newcommand{\fref}[1]{figure~\ref{#1}}
\newcommand{\sref}[1]{section~\ref{#1}}
\newcommand{\Aref}[1]{Appendix~\ref{#1}}
\newcommand{\aref}[1]{appendix~\ref{#1}}
\newcommand{\tref}[1]{table~\ref{#1}}
\newcommand{\rref}[1]{ref.~\cite{#1}}

\newcommand{\hypgeo}[2]{%
  \operatorname{%
    {\vphantom{\mathnormal{F}}}_{#1}%
    \kern-\scriptspace
    \mathnormal{F}_{#2}%
  }%
}
\makeatother

\begin{document}

\title{Quantum interference in gravitational particle production}

\author[a]{Edward Basso,}
\emailAdd{ebasso@wisc.edu}
\affiliation[a]{Department of Physics, The University of Wisconsin-Madison, 1150 University Avenue, Madison, WI 53706, U.S.A.}

\author[a]{Daniel J.\ H.\ Chung,}
\emailAdd{danielchung@wisc.edu}

\author[b]{Edward W.\ Kolb,}
\emailAdd{rocky.kolb@uchicago.edu}
\affiliation[b]{Kavli Institute for Cosmological Physics and Enrico Fermi Institute, University of Chicago, 5640 South Ellis Avenue, Chicago, IL 60637, U.S.A.}

\author[c]{and Andrew J.\ Long}
\emailAdd{andrewjlong@rice.edu}
\affiliation[c]{Department of Physics and Astronomy, Rice University, 6100 Main Street, Houston, TX 77005, U.S.A.}

\abstract{
Previous numerical investigations of gravitational particle production during the coherent oscillation period of inflation displayed unexplained fluctuations in the spectral density of the produced particles.  We argue that these features are due to the quantum interference of the coherent scattering reactions that produce the particles. We provide accurate analytic formulae to compute the particle production amplitude for a conformally-coupled scalar field, including the interference effect in the kinematic region where the production can be interpreted as inflaton scattering into scalar final states via graviton exchange. 
}

\date{\today}
\maketitle

\section{\label{sec:introduction}Introduction}

During the period of coherent oscillations of the inflaton field following the quasi-de Sitter (quasi-dS) phase of inflation \citep{Guth:1980zm,Starobinsky:1980te,Linde:1981mu,Albrecht:1982wi,Khlebnikov:1996mc,Prokopec:1996rr,Micha:2002ey}, particles (including dark matter candidates) may be produced via gravitationally-mediated nonthermal scattering processes in addition to the comparatively well-studied inflaton-decay and thermal-scattering processes \citep{Abbott:1982hn,Lyth:1996yj,Chung:1998zb,Chung:1998bt,Chung:2001cb,Dimopoulos:2006ms,Allahverdi:2010xz,Chung:2011ck,Ema:2015dka,Watanabe:2015eia,Graham:2015rva,Garny:2015sjg,Markkanen:2015xuw,Ema:2016hlw,Kannike:2016jfs,Hasegawa:2017hgd,Kolb:2017jvz,Tang:2017hvq,Garny:2017kha,Bernal:2018qlk,Alonso-Alvarez:2018tus,Fairbairn:2018bsw,Hashiba:2018iff,Garny:2018grs,Markkanen:2018gcw,Chung:2018ayg,Hashiba:2018tbu,Li:2019ves,Ema:2019yrd,Herring:2019hbe,Li:2020xwr,Chianese:2020yjo,Ema:2020ggo,Herring:2020cah,Ahmed:2020fhc,Chianese:2020khl,Kolb:2020fwh,Alexander:2020gmv,Redi:2020ffc,Garcia:2020wiy,Gross:2020zam,Ling:2021zlj,Mambrini:2021zpp,Dudas:2021njv,Antoniadis:2021jtg,Garcia:2021iag,Haque:2021mab,Clery:2021bwz,Haque:2022kez,Clery:2022wib,Garcia:2022vwm,Kaneta:2022gug,Mambrini:2022uol}.  Numerical investigations of gravitational particle production (GPP) employing the Bogoliubov approach have displayed unexplained oscillations as a function of the wavenumber $k$ in the final phase-space distribution $f_{\chi}(k)$ of the produced particles \citep{Giudice:1999am,Ema:2018ucl,Kolb:2021xfn}.  For example, figure~1 of \rref{Ema:2018ucl} showing the final phase-space density of dark-matter particle production in a hilltop inflationary model displays large oscillations that resemble numerical noise.  Similar large oscillations in the final phase-space density can be seen in the right-hand panel of figure~1 in \rref{Kolb:2021xfn} for GPP of the  helicity-1/2 component of a spin-3/2 Rarita-Schwinger field.

In this paper we explain these oscillatory features as the result of a quantum effect arising from an interference of different amplitudes, which are analogous to gravitationally-mediated nonthermal scattering processes,\footnote{We denote the inflaton as $\phi$ and the produced particle as $\chi$.  The produced particles, which may be dark matter candidates, are assumed to only interact gravitationally.  We sometimes denote $n\phi\rightarrow2\chi$ as the $n\phi$ process, for short.} $n\phi\rightarrow2\chi$ for $n \geq 1$. Typically the $2\phi$ process dominates nonthermal scattering production, but it has recently been pointed out \citep{Basso:2021whd} that the $n\phi$ processes with $n\neq2$ may also be important. Most of the effect comes from interference of $2\phi$ with the next leading amplitude, which is $3\phi$ if cubic interactions exist and $4\phi$ otherwise. We compute analytically the scattering contribution to the Bogoliubov amplitude including the interference terms, and find the results compare well with numerical computations. We also give a less technical semi-quantitative estimate of this interference amplitude based on a coherent scattering picture of a modified Boltzmann evolution. In this latter picture, the interference arises because the initial macroscopic inflaton scattering state can be viewed as a cold coherent superposition of $n\phi$ states, e.g., $c_{1}|\phi\phi\rangle+c_{2}|\phi\phi\phi\rangle$, such that the interference arises from $\left|c_{1}\langle\chi\chi|U|\phi\phi\rangle+c_{2}\langle\chi\chi|U|\phi\phi\phi\rangle\right|^{2}$ where $U$ schematically depicts a time evolution operator which is made more precise in this paper. Note that we write $n\phi\to2\chi$ to denote the net energy flow from the $\phi$ field to $\chi$ field, but this can be different from underlying S-matrix amplitudes. For example, $\phi\to2\chi$ has a contribution from the $\phi\phi\to\phi\chi\chi$ scattering process.

Although the quantum nature of the inflaton coherent-oscillation induced GPP has been known (e.g., see \citep{Kofman:1997yn}), the present paper extends the previous ideas to graviton-mediated scattering, and to our knowledge is the first to articulate clearly and to compute analytically  the quantum interference effects. It also clearly explains the previously unexplained ``noise'' in the particle production spectrum seen in the literature (see, e.g., \citep{Ema:2018ucl,Kolb:2021xfn}).  The application of a novel perturbative technique to solve the background inflaton dynamics is a technical highlight of this paper.

The order of presentation is as follows. In \sref{sec:ReviewOfGPP}, we give a brief review of the GPP computation using the Bogoliubov transform technique. In \sref{sec:Bogoliubov-computation}, we derive an analytic formula for the relevant Bogoliubov coefficient using a novel perturbation theory technique and a stationary phase approximation. The result is a sum of amplitudes analogous to $n\phi\rightarrow2\chi$, with \sref{subsec:Sample-analytic-amplitudes} presenting explicit results for $n\leq4$, and \sref{sec:discuss-interference} discussing the quantum interference between amplitudes. In \sref{sec:Numerical-comparison}, we compare the analytic results with numerical computations. In \sref{sec:A-Heuristic-derivation}, we interpret the interference as a novel contribution to the Boltzmann collision equation arising from the initial inflaton field being a macroscopic state described as a \emph{coherent superposition} of $n\phi$ states. We then conclude in \sref{sec:conclusion} with a summary and outlook.

The appendices contain some of the supporting technical details of this work. In \aref{sec:radiaangular} we describe the background field evolution in polar coordinates. In \aref{sec:Small--expansion}, we summarize the novel perturbation technique used to solve the inflaton dynamics with asymptotic series involving functions $\lambda(t)$ and $\theta(t)$ that describe slow and fast time scales, respectively. \Aref{sec:Phase--expansion} explains the technical details of evaluating the terms formally set up in the stationary phase computation in \sref{sec:Bogoliubov-computation} using this technique.  In \aref{sec:obtainingboltzmann}, we remind the reader how the statistical ensemble factor enters the usual collision integral of a Boltzmann equation in a manner that is in contrast with the picture of \sref{sec:A-Heuristic-derivation}. 

\section{\label{sec:ReviewOfGPP}Gravitational particle production}

Here, we focus on a background spacetime described by standard Einstein gravity with a spatially-flat Friedmann-Lema\^{i}tre-Robertson-Walker (FLRW) metric $ds^{2}=dt^{2}-a^{2}(t)|d\vec{x}|^{2} = a^2(\eta) \left[d\eta^2-|d\vec{x}|^{2}\right]$ where $d\eta=a^{-1}dt$ is conformal time.  The dominant energy-momentum tensor for the dynamics of the scale factor $a(t)$ comes from a minimally-coupled real scalar inflaton field $\phi$ with mass $m_\phi$ and a slow-roll inflationary potential $V(\phi)$.  We will assume that $m_{\phi}^{2}\equiv d^2V(\phi)/d\phi^2|_{\phi=v}\neq0$,  where $v$ is the minimum of $V(\phi)$ during the inflaton's coherent oscillation phase after the quasi-dS phase, and we will also assume that the nonlinearities in $V(\phi)$ can be captured as a Taylor expansion about $\phi=v$.  

The inflaton potential will be parameterized as 
\begin{align}
 V(\phi) & = 6M_P^{2}m_{\phi}^{2}\left[\frac{1}{2}\left(\frac{\phi-v}{\sqrt{6}M_P}\right)^{2}+\alpha_{3}\left(\frac{\phi-v}{\sqrt{6}M_P}\right)^{3}+\alpha_{4}\left(\frac{\phi-v}{\sqrt{6}M_P}\right)^{4}+\dots\right] \ , \label{eq:Vpert} \\
 \alpha_{n} & \equiv \frac{\left(\sqrt{6}M_P\right)^{n-2}}{m_{\phi}^{2}}\frac{1}{n!}\left.\frac{\partial^{n}V}{\partial\phi^{n}}\right|_{\phi=v} \ , \label{eq:alpha_n_def}
\end{align}
where $M_P=1/\sqrt{8\pi G}$ is the reduced Planck mass. When specific examples are needed, we will consider two inflaton models denoted by
\begin{align}
 V(\phi) & = \frac{1}{2} m_\phi^2\phi^2 & \  \mathrm{Quadratic} \ , \label{eq:QuadraticModelDef} \\
 V(\phi) & = \frac{m_\phi^2v^2}{72} \left[1-\frac{\phi^6}{v^6} \right]^2 & \mathrm{Hilltop} \ , \label{eq:HilltopModelDef}
\end{align}
where we take $v=0$ as the minimum for the Quadratic model and $v=M_P/2$ for the  Hilltop model.\footnote{For the sake of comparison, this is the same hilltop model considered by \rref{Ema:2018ucl}.}  Note that $\alpha_{3}=\alpha_{4}=0$ for the Quadratic potential while $\alpha_3=5\sqrt{6}$ and $\alpha_4=155$ for the Hilltop potential.

We augment the standard inflationary picture with a scalar ``spectator'' field $\tilde{\chi}$ whose action is given by
\begin{align}
 \Delta S = \int d\eta\, d^3\!x\  \frac{1}{2}\,  \left[(\partial_\eta\chi)^2-(\nabla\chi)^2-a^2\left(m_\chi^2+\frac{1}{6}(1-6\xi)R\right) \chi^2 \right] \ , \label{eq:chi_action}
\end{align}
where $\chi\equiv a\tilde{\chi}$ is the rescaled field, $m_\chi$ is the particle mass, and $R=-6\partial_\eta^2a/a^3$ is the Ricci scalar.  Following the usual procedure (e.g., \citep{Parker:1969au,Birrell:1982ix,Shtanov:1994ce,Kofman:1997yn}), we promote the scalar field to an operator $\hat{\chi}$ that satisfies the canonical equal-time commutation relations. The field operator is decomposed into mode functions $\chi_{{\bm{k}}}$ labeled by wavevector ${\bm k}$ as
\begin{align}
 \hat{\chi}({\bm{x}},\eta) = \int \frac{d^3\!\bm{k}}{2\pi^3} \left[\hat{\alpha}_{\bm{k}}\,\chi_{\bm{k}}(\eta)\,e^{i\bm{k}\cdot\bm{x}} + \hat{\alpha}_{\bm{k}}^\dagger\,\chi_{\bm{k}}^*(\eta)\,e^{-i\bm{k}\cdot\bm{x}}\right] \ ,
\end{align}
where the mode functions satisfy the normalization condition $\chi_{\bm{k}}\partial_\eta\chi_{\bm{k}}^* - \chi_{\bm{k}}^* \partial_\eta\chi_{\bm{k}}=i$, and the creation and annihilation operators satisfy the canonical commutation relations. Due to the action in \eref{eq:chi_action}, the mode equation is $\partial^2_\eta\chi_k+\omega_k^2\chi_k=0$, where
\begin{align}
 \omega_k^2=k^2+a^2\left(m_\chi^2+\frac{1}{6}(1-6\xi)R\right)
\end{align}
is the angular frequency of the $k^\mathrm{th}$ Fourier mode.\footnote{The mode functions will only depend on wavenumber $k=|\bm{k}|$ as the FLRW spacetime is isotropic.} The vacuum state $\left|0\rangle\right.$ is defined as $\hat{\alpha}_{\bm k}\left|0\rangle\right.=0$ for all ${\bm k}$, and particle creation is generated by the time-dependence of $\omega_k(\eta)$.  

While one can solve the mode equation directly given initial conditions, for our purposes we use the Bogoliubov parameterization. The mode functions are expressed as
\begin{align}
 \chi_k & = \frac{\alpha_k}{\sqrt{2\omega_k}}e^{-i\Omega_k} + \frac{\beta_k}{\sqrt{2\omega_k}}e^{+i\Omega_k} \ , \label{eq:BC}
\end{align}
\begin{align}
 \Omega_k(t) & \equiv \int_{t_i}^{t} dt'\sqrt{\frac{k^2}{a^2(t')}+m_\chi^2+\frac{1}{6}(1-6\xi)R(t')} \equiv \int_{t_i}^{t}dt'\,E_k(t')\label{eq:omega_def} \ ,
\end{align}
where $\alpha_k$ and $\beta_k$ are the Bogoliubov coefficients, which decompose the mode function into positive and negative-frequency components, respectively. 
The nearly-adiabatic conditions in the far past motivates the Bunch-Davies initial condition such that $\alpha_k=1$ and $\beta_k=0$ at initial time $t=t_i$. 
In the evolution of $\chi_k$ from the initial negative-frequency solution, a positive-frequency component may appear, signaling particle creation.  
In the far late-time, the number density of produced particles is given by
\begin{align}
 n_\chi(t) a^3(t) = \int\frac{d^3k}{(2\pi)^3} \ f_\chi(k,t) \ , \label{eq:f_chi_def}
\end{align}
where $f_\chi(k,t) \equiv \left|\beta_{k}(t)\right|^{2}$ denotes the produced $\chi$-particle phase-space density. We therefore seek a solution for $\beta_k$ to compute GPP. The time-evolution of the Bogoliubov coefficients $\alpha_k$ and $\beta_k$ can be written as\footnote{We will use the notation $\dot{x}\equiv dx/dt$ throughout this paper.}
\begin{align}
 \dot{\alpha}_{k}(t) & =\widetilde{\mathcal{N}}_{k}(t)\,\beta_{k}(t)\,e^{+2i\Omega_{k}(t)} \\
 \dot{\beta}_{k}(t) & =\widetilde{\mathcal{N}}_{k}(t)\,\alpha_{k}(t)\,e^{-2i\Omega_{k}(t)}\label{eq:betakdoteq}
\end{align}
as is done for example in \citep{Brezin:1970xf,Kofman:1997yn,Chung:1998bt}.  For the case of a scalar $\chi$ field, we use the definition
\begin{align}
 \widetilde{\mathcal{N}}_{k} & \equiv \frac{\dot{\omega}_{k}}{2\omega_{k}}  = \frac{1}{2}\frac{H m_\chi^2 + \frac{1}{6}(1-6\xi)(HR+\frac{1}{2}\dot{R})}{k^2 / a^2+m_\chi^2+\frac{1}{6}(1-6\xi)R}   \label{eq:tildeN} \ ,
\end{align} 
with $\Omega_k$ defined in \eref{eq:omega_def}. The background evolution (assumed driven by the dynamics of the inflaton) enters the determination of $\widetilde{\mathcal{N}}_{k}$ through $a$, $H$, $R$, and $\dot{R}$, while the spectator field enters through $k$ and $m_\chi$. Setting $\alpha_{k}\approx1$ in \eref{eq:betakdoteq}, we write
\begin{equation}
 \boxed{\beta_{k}=\int_{t_{i}}^{t_{f}}dt\,\widetilde{\mathcal{N}}_{k}(t) e^{-2i\Omega_{k}(t)} \ ,}\label{eq:beta_integral}
\end{equation}
which is valid for $|\beta_{k}|\ll1$. This important integral expression is the staring point for the main results of this paper.

\section{\label{sec:Bogoliubov-computation}Novel computation of the Bogoliubov coefficient}

In situations where there are coherent oscillations of the inflaton, \textit{some} contributions to GPP can be interpreted as coming from scattering of the inflaton quanta into $\chi$s via graviton exchange \citep{Ema:2015dka,Ema:2016hlw,Ema:2018ucl,Kaneta:2022gug}.  Below, we explain a novel computation of this coefficient that gives not only the amplitudes of $n\phi \to 2\chi$ during the inflaton coherent oscillations, but also the \emph{interference} between amplitudes with different $n\geq 1$.

In the scenario with coherent oscillations of the inflaton field after the quasi-dS phase, the Hubble expansion rate has two broad classes of components: intuitively, $H=H_{\mathrm{slow}}+H_{\mathrm{fast}}$, where the oscillatory fast component is smaller in amplitude but varies on a larger frequency scale compared to the monotonically decreasing slow component. The respective scales are $\frac{d}{dt}\ln H_\mathrm{slow}\sim H$ and $\frac{d}{dt}\ln H_\mathrm{fast}\sim m_{\phi}$, and therefore $H<m_\phi$ is required for a meaningful distinction. This condition holds in the oscillatory era for most single field inflationary models. This decomposition is made precise using the formalism summarized in \aref{sec:Small--expansion}. This approach differs from that of \rref{Basso:2021whd}, where time was partitioned into bins of size $H_{\mathrm{slow}}^{-1/2}m_{\phi}^{-1/2}$ to find the $\dot{\phi}$ behavior applicable to each of those bins. That approach is suitable for the particle production computation without the interference, but to find the interference, we need to keep track of the time phase of the Bogoliubov coefficients across different time bins $\Delta t$. The approach we will take below is to use a novel method of expanding the background field evolution in $H_{\mathrm{slow}}/m_{\phi}$.

This formalism introduces functions $\lambda(t)$ and $\theta(t)$ that partition the slow-time and fast-time dependence of a general quasi-periodic function of time such as $\widetilde{\mathcal{N}}_k(t)$. The monotonically decreasing slow-time variable $\lambda$ can be thought of as 
\begin{equation}
 \lambda(t)\sim \frac{H_{\mathrm{slow}}(t)}{m_\phi}\sim\frac{H_{e}/m_\phi}{1+\frac{3}{2}H_{e}(t-t_{e})}\sim\frac{2}{3m_\phi t}\label{eq:lambdacorrespondence} \ ,
\end{equation}
where all quantities with a subscript-$e$ index will refer to its value at the time $t_e$ when the quasi-dS era ends, also referred to as the end of inflation. For the two inflation models we will consider, Quadratic and Hilltop,  $H_e/m_\phi \simeq 0.5$ and $0.03$, respectively (see \tref{tab:Hbar_Xibar_abar_table}), and therefore $\lambda(t)$ is less than unity for $t \geq t_e$. The fast-time variable $\theta$ can be thought of as a diffeomorphism of time to a monotonically increasing phase function such that 
\begin{align}
\theta(t)\sim m_\phi t\ .
\end{align} 
In short, $\lambda$ and $\theta$ describe time scales of $H^{-1}$ and $m_\phi^{-1}$, respectively. This partitioning is the basis for the novel perturbation technique appearing in \rref{preplamtheta}, which allows us to resum secular effects and track $\beta_{k}(t)$ accurately for a long time (a time much longer than $m_{\chi}^{-1}$).  For example, the Hubble expansion rate is expanded systematically as
\begin{equation}
 H=m_{\phi}\lambda\left(1+\sum_{\ell=1}^{\infty}h_{\ell}(\theta)\lambda^{\ell}\right) \ , \label{eq:H_expansion}
\end{equation}
where $h_{\ell}(\theta)$ contains the fast-time behavior as a sum of sinusoids that depend on \emph{integer multiples of $\theta$}. The higher-integer frequency components become increasingly negligible as they generally come with higher powers of $\lambda(t)$. This accurate tracking for a long time is useful for capturing our sought-after interference effects, which develop on a time scale $\Delta t \gg H^{-1}\gg m_{\chi}^{-1}$ for the $k/a_{e} > m_{\phi}$ modes. The time evolution of these functions is defined by 
\begin{align}
  \dot{\lambda} & =-\frac{3}{2}m_\phi\lambda^{2}(1+\mathcal{O}(\lambda^2)) \ , \label{eq:lam_dot_leading_behavior} \\ 
  \dot{\theta} & = m_\phi (1+\mathcal{O}(\lambda^2)) \ , \label{eq:theta_dot_leading_behavior}
\end{align}
where the initial conditions and details of the $\mathcal{O}(\lambda^2)$ corrections are irrelevant for the arguments of this section. The $\lambda,\theta$ decomposition is explained in further detail in \aref{sec:Small--expansion}.

To describe the interference effects, we write $\beta_k$ as a sum over $\beta_k^{(n\to2)}$ contributions by partitioning the time dependence of the integrand from \eref{eq:beta_integral} into $\lambda(t)$ and $\theta(t)$, where an exact definition of $\beta_k^{(n\to2)}$ is presented below. In summary, the $n\to2$ label alludes to the $n\phi\to2\chi$ scattering process, with the integer $n$ denoting the frequency that stems from an oscillatory dependence on $n\theta$. As noted below \eref{eq:H_expansion} and shown in \sref{subsec:Sample-analytic-amplitudes}, larger $n$ are increasingly suppressed, and therefore $\beta_k$ is well approximated by the first few terms of this sum. This property motivates the use of this formalism, and is analogous to the suppression of higher particle number processes in perturbative QFT.

We now present definitions used in the specification and computation of $\beta_k^{(n\to2)}$. Given a general quantity $X(\lambda,\theta)$, we define the convention 
\begin{equation}
 X(\lambda,\theta)=\sum_{n=-\infty}^{\infty}X^{(n)}(\lambda)e^{in\theta} \ , \label{eq:nth_decomposition}
\end{equation}
\begin{equation}
 X^{(n)}(\lambda)\equiv\frac{1}{2\pi}\int_{0}^{2\pi}X(\lambda,\theta)e^{-in\theta}d\theta \label{eq:nth} \ ,
\end{equation}
where $X^{(n)}(\lambda)$ is the $n^\mathrm{th}$-frequency component of $X$. 
The slow-time component $X_s$ is defined as the $n=0$ term, which does not contain any fast-time information, i.e., no $\theta$-dependence, and the fast component $X_f$ is defined as the remainder (i.e., the sum of the $n\neq0$ terms). As an example, it will be useful to separate the phase-factor $\Omega_k$ into a slow component $\Omega_{s,k}(\lambda)\equiv \Omega^{(0)}_{k}(\lambda)$ and a fast component $\Omega_{f,k}(\lambda,\theta) \equiv \Omega_k(\lambda,\theta) - \Omega_{s,k}(\lambda)$. 

We define $\beta_k^{(n\to2)}$ for the resonant scattering situations of current interest such that  
\begin{align}
	\beta_{k} 
	= \sum_{n=1}^{\infty}\beta_{k}^{(n\rightarrow2)} 
	\qquad \text{where} & \qquad 
	\beta_k^{(n\rightarrow2)}\equiv\int_{t_{i}}^{t_{f}}dt\,\mathcal{N}_{k}^{(n)}(\lambda)\,e^{i(n\theta-2\Omega_{\mathrm{s},k}(\lambda))} \ , \label{eq:Bkn} 
 \end{align}
and $\mathcal{N}_k\equiv \widetilde{\mathcal{N}}_k e^{-2 i \Omega_{f,k}}$, with the transformation of $t$ to $\lambda,\theta$ dependence as well as the expression of $\widetilde{\mathcal{N}}_k(\lambda,\theta)$ described in appendices~\ref{sec:Small--expansion}~and~\ref{sec:Phase--expansion}.\footnote{The integral of \eref{eq:Bkn} for $n\leq0$ is exponentially suppressed for resonant scattering, i.e., $k \gtrsim m_\phi a_e$, which is why the sum of \eref{eq:Bkn} starts from $n=1$.} The exponential in \eref{eq:Bkn} hints at the relation to the amplitude for $n\phi\to 2\chi$ as its phase is stationary when $n m_\phi \approx 2 \sqrt{k^2/a^2 + m_\chi^2}$, which corresponds to the energy condition of $n$ inflatons at rest annihilating to produce two $\chi$ particles with momentum $k/a$. This correspondence to scattering motivates referring to $\beta_k^{(n\to2)}$ as the $n\to2$ resonance component, and suggests evaluation using the stationary phase approximation.

We compute the $n\rightarrow2$ resonance component of the Bogoliubov coefficient using the stationary phase approximation, which will ultimately lead to an expansion in powers of $k^{-3/2}$. For the purposes of explaining the computation, we write
\begin{equation}
	\Psi_{k}^{(n)}(t) \equiv in\theta(t) - 2i\Omega_{s,k}(\lambda(t)) + \log \mathcal{N}_k^{(n)}(\lambda(t))  \label{eq:phasePsi}
\end{equation}
as the total (complex-valued) phase for the $n\rightarrow2$ resonance. The phase is stationary when $\dot{\Psi}_{k}^{(n)}(\overline{t}_{k}^{(n)})\equiv0$, where we call $\overline{t}_{k}^{(n)}$ as the resonance time,\footnote{While more than one stationary point may exist, we assume there is a single dominant point.} which will usually have a small imaginary component due to the complex nature of the phase. The phase is expanded as 
\begin{align}
  \beta_{k}^{(n\rightarrow2)}&
  =e^{\Psi_{k}^{(n)}(\overline{t}_{k}^{(n)})}\int_{t_{i}}^{t_{f}}dt\,\exp\left(\sum_{\ell=2}^{\infty}\frac{1}{\ell!}\,\partial_{t}^{\ell}\Psi_{k}^{(n)}(\overline{t}_{k}^{(n)})\,(t-\overline{t}_{k}^{(n)})^{\ell}\right) \ , 
  \label{eq:beforedeform}
\end{align}
which is the starting point of the stationary phase approximation. We now  define a new variable $z$ such that the quadratic term in the exponential becomes $-\frac{1}{2}z^{2}$, and evaluate using an expansion of Gaussian integrals as 
\begin{align}
 \beta_{k}^{(n\rightarrow2)}&=\frac{e^{\Psi_{k}^{(n)}(\overline{t}_{k}^{(n)})}}{\sqrt{-\ddot{\Psi}_{k}^{(n)}(\overline{t}_{k}^{(n)})}}\int_{-\infty}^{+\infty}dz\,e^{-\frac{1}{2}z^{2}}\,\exp\left(\sum_{\ell=3}^{\infty}\frac{\partial_{t}^{\ell}\Psi_{k}^{(n)}(\overline{t}_{k}^{(n)})}{\ell![-\ddot{\Psi}_{k}^{(n)}(\overline{t}_{k}^{(n)})]^{\ell/2}}z^{\ell}\right)
 \label{eq:Taylor_series_stationary_phase} \\
 & =\frac{e^{\Psi_{k}^{(n)}(\overline{t}_{k}^{(n)})}\sqrt{2\pi}}{\sqrt{-\ddot{\Psi}_{k}^{(n)}(\overline{t}_{k}^{(n)})}}\left(1-\frac{5[\partial_{t}^{3}\Psi_{k}^{(n)}(\overline{t}_{k}^{(n)})]^{2}}{24[\ddot{\Psi}_{k}^{(n)}(\overline{t}_{k}^{(n)})]^{3}}+\frac{\partial_{t}^{4}\Psi_{k}^{(n)}(\overline{t}_{k}^{(n)})}{8[\ddot{\Psi}_{k}^{(n)}(\overline{t}_{k}^{(n)})]^{2}}+\dots\right)\ ,
 \label{eq:stationaryphaseresult}
\end{align}
where the coefficients of the higher powers of $z$ are treated as increasingly negligible, an assumption that will be justified shortly. In going from \eref{eq:beforedeform} to \eref{eq:Taylor_series_stationary_phase}, we moved the contour into the complex plane in addition to changing the integration limits. In particular, the contour was rotated by approximately 45 degrees as $\ddot{\Psi}_k^{(n)}$ at resonance is dominated by its imaginary component due to the first two terms of \eref{eq:phasePsi}.

We need an expansion parameter to truncate the expansion of \eref{eq:stationaryphaseresult}, which we will see is proportional to $k^{-3/2}$ by the following argument. Since $\dot{\lambda}\sim\lambda^2$ and the derivatives of the phase depend on $\lambda$, we know that $\partial_t^{\ell}\Psi_k^{(n)}\sim\lambda^{\ell-1}$ and therefore the $z^\ell$ coefficient in \eref{eq:Taylor_series_stationary_phase} scales as $\lambda^{\ell/2 -1}$, which is suppressed for $\ell\geq3$. Hence, the next step is to evaluate $\lambda(\bar{t}_{k}^{(n)})$. The phase is stationary when $n m_\phi\approx\sqrt{k^2/a^2+m_\chi^2}$ for $\mathcal{N}_{k}^{(n)}$ that is non-singular in $\lambda$. The scale factor satisfies $a=\underline{a}\lambda^{-2/3}(1+\mathcal{O}(\lambda))$, where the constant of integration $\underline{a}\sim a_e H_e^{2/3}/m_\phi^{2/3}$ was determined by definition in \eref{eq:a_underline_def}. Therefore, we estimate
\begin{equation}
  \lambda(\overline{t}_{k}^{(n)})\approx\left(\frac{k/\underline{a}}{\sqrt{\tfrac{n^{2}}{4}m_{\phi}^{2}-m_{\chi}^{2}}}\right)^{-3/2}\sim\frac{H_{e}}{m_{\phi}}\left(\frac{k/a_e}{\sqrt{\frac{n^{2}}{4}m_{\phi}^{2}-m_{\chi}^{2}}}\right)^{-3/2} \ , \label{eq:lamkhrelation}
 \end{equation}
 which is a small number for all $k\gtrsim H_{e}^{2/3}m_{\phi}^{1/3}a_{e}$ for $n = \mathcal{O}(1)$. Since our present computation is focusing on scattering of particle-like modes at the end of the quasi-dS era, $\lambda(\overline{t}_{k}^{(n)})$ is small for all the modes of our present interest and thus is na\"{i}vely a useful expansion parameter. This statement will be made more sharp in \sref{sec:Numerical-comparison}.

We would like to solve for $\lambda(\overline{t}_{k}^{(n)})$ itself as a function of $k$. Hence, we define a separate expansion parameter for the stationary phase approximation as
\begin{equation}
  \varepsilon\,\kappa_{n}^{-3/2}\equiv\left(\frac{k/\underline{a}}{\sqrt{\tfrac{n^{2}}{4}m_{\phi}^{2}-m_{\chi}^{2}}}\right)^{-3/2} \ , \label{eq:epskappadef}
\end{equation}
where $\varepsilon$ is a bookkeeping parameter inspired by the smallness of \eref{eq:lamkhrelation}. We can parameterize a perturbation series solution of $\lambda(\overline{t}_{k}^{(n)})$ in powers of $\varepsilon\, \kappa_{n}^{-3/2}$ as 
\begin{equation}
\lambda(\overline{t}_{k}^{(n)})=\varepsilon\,\kappa_{n}^{-3/2}\left(1+\sum_{j=1}^{\infty}r_{k,j}^{(n)}\left(\varepsilon\,\kappa_{n}^{-3/2}\right)^{j}\right) \label{eq:lambdakappa}
\end{equation}
where the constant coefficients $r_{k,j}^{(n)}$ are determined by solving the stationary phase condition $\dot{\Psi}_{k}^{(n)}(\overline{t}_{k}^{(n)})=0$ at each order in $\varepsilon$. When using the replacement of \eref{eq:lambdakappa}, it is important to write $k$ in terms of $\varepsilon^{-2/3}\kappa_n$ using \eref{eq:epskappadef} to cancel out the fractional powers of $\varepsilon$ that appear due to $a\sim\lambda^{-2/3}$. This is equivalent to assuming $k$ and $a m_\phi$ have the same magnitude at resonance. Afterwards, we can use our solution to evaluate the phase-derivative coefficients appearing in \eref{eq:stationaryphaseresult}. Some of the technical details of this computation are given in \aref{sec:Phase--expansion}.

\section{\label{subsec:Sample-analytic-amplitudes}Analytic formulas for the Bogoliubov coefficient}

In this section, we explicitly list the analytic amplitudes $\beta_{k}^{(n\rightarrow2)}$ for $n\in\{1,2,3,4\}$ solved by the procedure described above, with the $k$-dependence expressed as an expansion in $\kappa_n^{-3/2}$ as defined by \eref{eq:epskappadef}. We choose a conformally-coupled ($\xi=1/6$) scalar $\chi$ field because of the relative simplicity of the source of nonadiabaticity. To make the interference phase more manifest, we express our results as 
\begin{align}\label{eq:betak_to_Ak_Phik}
	\beta_{k}^{(n\rightarrow2)}=\mathcal{A}_{k}^{(n\rightarrow2)}e^{i\Phi_{k}^{(n\rightarrow2)}} \ .
\end{align}
Up to a global phase that is independent of $n$, which therefore affects neither the interference nor the magnitude of $\beta_k$, the leading terms for the phase can be written as
\begin{equation}
 \Phi_{k,\mathrm{leading}}^{(n\rightarrow2)}=\frac{2}{3}\kappa_{n}^{3/2}\left(n-2r_{\chi}\hypgeo{2}{1}\left(-\tfrac{3}{4},-\tfrac{1}{2};\tfrac{1}{4};1-\tfrac{n^{2}}{4r_{\chi}^{2}}\right)\right)+n\left(\underline{\Xi}-\frac{2m_{\phi}}{3\underline{H}}\right) \ , \label{eq:leadingPhi}
\end{equation}
where $r_{\chi}\equiv m_{\chi}/m_{\phi}$, $\underline{H}$ and $\underline{\Xi}$ are boundary conditions defined in \erefs{eq:H_underline_def}{eq:Xi_underline_def}, respectively, and $\hypgeo{2}{1}$ is the hypergeometric function.  We will give a physical interpretation of this leading order phase in \sref{sec:A-Heuristic-derivation}. If we define $\Delta\Phi_k^{(n\to2)}\equiv\Phi_k^{(n\to2)}-\Phi_{k,\mathrm{leading}}^{(n\to2)}$ and $r_\chi=m_\chi/m_\phi$, then we can write our results as 
\begin{subequations}\label{eq:Ak}
\begin{align}
 \mathcal{A}_{k}^{(1\rightarrow2)} & =-\kappa_1^{-15/4} \, 3 \alpha_{3}\sqrt{\frac{-\frac{i}{2}\pi}{\frac{1}{4}-r_{\chi}^{2}}} \, r_\chi^2\left(1+\mathcal{O}(\kappa_{1}^{-3})\right) \ , \\
 \mathcal{A}_{k}^{(2\rightarrow2)} & =\kappa_2^{-9/4} \frac{3}{16}\sqrt{\frac{-i\pi}{1-r_{\chi}^{2}}} \, r_\chi^2  \left(1+  \frac{x_{0}+x_{1}r_{\chi}^{2}+x_{2}r_{\chi}^{4}-416r_{\chi}^{6}+384r_{\chi}^{8}}{1024(1-r_{\chi}^{2})^{2}} \kappa_{2}^{-3} +\mathcal{O}(\kappa_{2}^{-6})\right) \ , \\ \label{eq:2to2expl}
\mathcal{A}_{k}^{(3\rightarrow2)} & = \kappa_3^{-15/4}\, \frac{\alpha_{3}}{9}\sqrt{\frac{-\frac{3}{2}i\pi}{\frac{9}{4}-r_{\chi}^{2}}} \,  r_\chi^2\left(1+\mathcal{O}(\kappa_{3}^{-3})\right) \ ,\\
\mathcal{A}_{k}^{(4\rightarrow2)} & = \kappa_4^{-21/4}\, \frac{3\left(-21+68\alpha_{3}^{2}+24\alpha_{4}+12r_{\chi}^{2}\right)}{4096}\sqrt{\frac{-2i\pi}{4-r_{\chi}^{2}}} \, r_\chi^2 \left(1+\mathcal{O}(\kappa_{4}^{-3})\right) \ , 
\end{align}
\end{subequations}
\begin{subequations} \label{eq:Phik}
\begin{align}
 \Delta\Phi_{k}^{(1\rightarrow2)} & =\kappa_1^{-3/2} \left(\frac{y_{0}^{(1)}+y_{1}^{(1)}r_{\chi}^{2}-1280r_{\chi}^{4}}{480\left(1-4r_{\chi}^{2}\right)}+z^{(1)}+\mathcal{O}(\kappa_{1}^{-3}) \right) \ , \\
\Delta\Phi_{k}^{(2\rightarrow2)} & = \kappa_2^{-3/2}\left(\frac{y_{0}^{(2)}+y_{1}^{(2)}r_{\chi}^{2}-80r_{\chi}^{4}}{960\left(1-r_{\chi}^{2}\right)}+z^{(2)}+\mathcal{O}(\kappa_{2}^{-3}) \right) \ , \\  
\Delta\Phi_{k}^{(3\rightarrow2)} & =\kappa_3^{-3/2}\, \left(\frac{y_{0}^{(3)}+y_{1}^{(3)}r_{\chi}^{2}-1280r_{\chi}^{4}}{12960\left(9-4r_{\chi}^{2}\right)}+z^{(3)} + \mathcal{O}(\kappa_3^{-3}) \right) \ , \\
\Delta\Phi_{k}^{(4\rightarrow2)} & =\kappa_4^{-3/2} \, \left(\frac{y_{0}^{(4)}+y_{1}^{(4)}r_{\chi}^{2}+y_{2}^{(4)}r_{\chi}^{4}+2588r_{\chi}^{6}}{960\left(4-r_{\chi}^{2}\right)\left(-21+68\alpha_{3}^{2}+24\alpha_{4}+12r_{\chi}^{2}\right)}+z^{(4)}+\mathcal{O}(\kappa_{4}^{-3}) \right) \ ,
\end{align}
\end{subequations}
where $x_{i},y_{i}^{(n)},z^{(n)}$ are merely notational variables to allow a visually manageable display of the results, with their explicit values given in \aref{sec:xyz_coefficients}. Given the generalized nature of $\alpha_n$ as defined in \eref{eq:alpha_n_def}, \emph{these results are applicable to any inflationary potential which can expanded as a polynomial with a positive quadratic term at its minimum}. Since $\mathcal{A}_k^{(2\to2)}$ will be the dominant term, we have shown it to higher order in the expansion. Note that the higher order terms in these formulas are organized by time reversal symmetry, and therefore have relative corrections that go as $\varepsilon^2 \kappa_n^{-3}$ instead of $\varepsilon \kappa_n^{-3/2}$.

\begin{table}
  \begin{centering}
   \begin{tabular}{>{\raggedright}p{1.8cm}>{\centering}p{1.8cm}>{\centering}p{1.8cm}>{\centering}p{1.8cm}>{\centering}p{1.8cm}>{\centering}p{1.8cm}>{\centering\arraybackslash}p{1.8cm}}
     Model & $m_{\chi}/H_{e}$ & $m_{\phi}/H_{e}$ & $\Xi_{e}$ & $m_{\phi}/\underline{H}$ & $\underline{\Xi}$ & $\underline{a}/a_{e}$\tabularnewline
     \hline 
     Quadratic & $0.1$ & $1.981$ & $0.615$ & $2.748$ & $-0.138$ & $0.5662$\tabularnewline
     Hilltop & $10$ & $31.31$ & $-2.526$ & $26.90$ & $-9.755$ & $0.1041$\tabularnewline
   \end{tabular}
  \par\end{centering}
 \caption{\label{tab:Hbar_Xibar_abar_table}Values in the Quadratic and Hilltop models of inflation defined by \erefs{eq:QuadraticModelDef}{eq:HilltopModelDef}, respectively. Note that dividing $\underline{a}/a_e$ by $(H_e/m_\phi)^{2/3}$ gives $0.893$ and $1.034$ for the respective models, which are $\mathcal{O}(1)$, as expected based on the definition of $\underline{a}$ in \eref{eq:a_underline_def}.}
\end{table}

The boundary conditions for the background functions $\{\phi(t),a(t)\}$ are given by 3 integration constants and are needed for specifying analytic formulas for $\beta_{k}^{(n\rightarrow2)}$. These can be chosen to be $\underline{H}$, $\underline{\Xi}$, and $\underline{a}$ defined by limits as $t\to\infty$ in \erefss{eq:H_underline_def}{eq:Xi_underline_def}{eq:a_underline_def}, respectively. This method of writing the boundary conditions allows a cleaner set of analytic expressions. Numerical solutions were used to obtain the values found in \tref{tab:Hbar_Xibar_abar_table} for two inflationary models of interest, and this was done mostly for accuracy when comparing the analytic $\beta_{k}$ with numerically computed $\beta_{k}$. Analytic expressions can be obtained for these integration constants as an expansion in $H_e/m_\phi$ for standard slow-roll inflationary scenarios entering the coherent oscillations period. For example, \erefss{eq:H_underline_analytic_estimate}{eq:Xi_underline_analytic_estimate}{eq:a_underline_analytic_estimate} only require conditions at $t_e$ and give comparable values to those found in \tref{tab:Hbar_Xibar_abar_table}: $m_\phi/\underline{H}\approx 2.799$, $\underline{\Xi}\approx -0.078$, and $\underline{a}/a_e\approx0.5571$ for the Quadratic model, and $m_\phi/\underline{H}\approx 27.28$, $\underline{\Xi}\approx -9.144$, and $\underline{a}/a_e\approx0.1038$ for the Hilltop model.

The calculations were done without choosing a particular scheme, a type of gauge choice concept that is particular to our computational formalism briefly described in \aref{sec:Small--expansion}. Instead, the scheme dependence was kept general throughout and completely cancelled out in the final result. Checking scheme independence of observables was a robust tool to verify different steps of the calculation. Another feature to note is that all amplitudes vanish as they should when $m_{\chi}/m_{\phi}\rightarrow0$ since we are considering the conformally-coupled case. A related feature is that the leading phase $\Phi_{k,\mathrm{leading}}^{(n\rightarrow2)}$ diverges in the limit that $m_{\chi}/m_{\phi}\rightarrow0$.

Let's now compare the squared amplitudes between the current computation and an earlier work by some of the present authors \citep{Basso:2021whd}. From the latter, we have the estimate
\begin{align}
 f_{\chi}(k) & =\frac{9\pi}{256}\frac{\tilde{H}_{\mathrm{end}}^{3}}{m_{\phi}^{3}\bigl(1-m_{\chi}^{2}/m_{\phi}^{2}\bigr)}\left(\frac{m_{\chi}}{m_{\phi}}\right)^{4}\left(\frac{k/a_{\mathrm{end}}}{\sqrt{m_{\phi}^{2}-m_{\chi}^{2}}}\right)^{-9/2} \ ,
\end{align}
where the definition of $\tilde{H}_\mathrm{end}$ and the above equation for $f_\chi$ are given by eqs.\,(8.13) and (8.17) of \rref{Basso:2021whd}, respectively. This can be compared to our \eref{eq:2to2expl}.  
For the Hilltop model of \eref{eq:HilltopModelDef}, the leading expressions differ by a factor of 
\begin{equation}
\frac{\left|\mathcal{A}_{k}^{(2\rightarrow2)}\right|^{2}}{f_{\chi}(k,t)}\approx\frac{m_\phi^3 }{\tilde{H}_\mathrm{end}^3}\frac{\underline{a}^{9/2}}{a_\mathrm{end}^{9/2}}\approx\frac{(30.41 H_\mathrm{end})^3(0.1041a_e)^{9/2}}{(1.843H_\mathrm{end}^3)(0.875a_e)^{9/2}} = 1.052
\end{equation}
where the value of $\underline{a}$ is found in \tref{tab:Hbar_Xibar_abar_table}, and the value of $\tilde{H}_{\mathrm{end}}$ for this Hilltop model is given by eq.\,(8.27) of \rref{Basso:2021whd}. The difference between $a_e$ and $a_\mathrm{end}$ (also $H_e$ and $H_\mathrm{end}$) is a result of different definitions for the end of inflation. This ratio can be used as an estimate of corrections that this paper represents to the computations of \rref{Basso:2021whd} as far as the non-interference piece is concerned. 

\section{\label{sec:discuss-interference}Discussion of the interference}

Now, let's consider the interferences arising from the results of \sref{subsec:Sample-analytic-amplitudes}. To focus the discussion
to the physically most significant case, consider the interference between $2\rightarrow2$ and $3\rightarrow2$ amplitudes: 
\begin{equation}
\left|\mathcal{A}_{k}^{(2\rightarrow2)}e^{i\Phi_{k}^{(2\rightarrow2)}}+\mathcal{A}_{k}^{(3\rightarrow2)}e^{i\Phi_{k}^{(3\rightarrow2)}}\right|\ni2\mathrm{Re}\left\{ \mathcal{A}_{k}^{(2\rightarrow2)}\mathcal{A}_{k}^{*(3\rightarrow2)}e^{i\left[\Phi_{k}^{(2\rightarrow2)}-\Phi_{k}^{(3\rightarrow2)}\right]}\right\} \ .
\end{equation}
Since $\mathcal{A}_{k}^{(2\rightarrow2)}\mathcal{A}_{k}^{*(3\rightarrow2)}$
is real, the interference phase between these two processes comes from 
\begin{align}
\Phi_{k}^{(2\rightarrow2)}-\Phi_{k}^{(3\rightarrow2)} & =\frac{2m_{\phi}}{3\underline{H}}-\underline{\Xi}+\frac{2}{3}\left(2\kappa_{2}^{3/2}-3\kappa_{3}^{3/2}\right)-\frac{4r_{\chi}}{3}\kappa_{2}^{3/2}\hypgeo{2}{1}\left(-\tfrac{3}{4},-\tfrac{1}{2};\tfrac{1}{4};1-\tfrac{1}{r_{\chi}^{2}}\right)\nonumber \\
 & +\frac{4}{3}\frac{m_{\chi}}{m_{\phi}}\kappa_{3}^{3/2}\hypgeo{2}{1}\left(-\tfrac{3}{4},-\tfrac{1}{2};\tfrac{1}{4};1-\tfrac{9}{4r_{\chi}^{2}}\right)+\Delta+\mathcal{O}(\varepsilon^3)\label{eq:interference-phase} \ ,
\end{align}
where $r_\chi=m_\chi/m_\phi$, and $\Delta$ is defined as
\begin{align}
\Delta&\equiv\frac{1}{\kappa_{2}^{3/2}}\left(\frac{y_{0}^{(2)}+y_{1}^{(2)}r_{\chi}^{2}-80m_{\chi}^{4}}{960\left(1-r_{\chi}^{2}\right)}+z^{(2)}\right)-\frac{1}{\kappa_{3}^{3/2}}\left(\frac{y_{0}^{(3)}+y_{1}^{(3)}r_{\chi}^{2}-1280r_{\chi}^{4}}{12960\left(9-4r_{\chi}^{2}\right)}+z^{(3)}\right) \ , 
\end{align}
with $y_{m}^{(n)}$ as numerical coefficients that depend only on the inflaton potential interaction strengths $\alpha_{3}$ and $\alpha_{4}$, as can be seen in \aref{sec:xyz_coefficients}. The term proportional to $2\kappa_{2}^{3/2}-3\kappa_{3}^{3/2}$ comes from $2\theta(\bar{t}_{k}^{(2)})-3\theta(\bar{t}_{k}^{(3)})$, and each of these terms with the respective coefficients are effectively a rewriting of the resonance times. The hypergeometric functions correspond to the $2\Omega_{k}$ phases appearing in \eref{eq:betakdoteq} evaluated at the respective resonance times. Equation (\ref{eq:interference-phase}) is one of the main analytic results of this present work.

The $\Delta$ term contains the leading higher-$\lambda$ power correction to the leading stationary-phase result. This contains the nontrivial corrections to the phases coming from the cubic and quartic interaction terms of the inflaton potential: i.e., it depends on $\alpha_{3,4}$.   It vanishes in the large $k/m_{\phi}$ limit because this is just the property of an asymptotic expansion through the stationary phase method.

In \sref{sec:A-Heuristic-derivation}, we will discuss how the phases can be interpreted in terms of phases accumulating through the Hamiltonian energy driven time evolution. In this intuitive picture, for a given time interval, the inflaton background field self-interaction and self-gravitational interaction change the accumulated phase of the inflaton interpreted as a collection of one-particle states because of the change in the effective free propagator Hamiltonian energy. For example, in the parameter region of $\{m_{\phi}\gg m_{\chi},\alpha_{3}=0,m_{\phi}\gg H\}$, one can easily check that $\Delta$ increases as expected from the intuition that the steepening of the potential by the quartic potential contribution increases the effective oscillation mass. The  $\kappa_{2}^{-3}$ correction term in $|\mathcal{A}_{k}^{(2\rightarrow2)}|^{2}$ of \eref{eq:2to2expl} increases with $m_{\chi}$ increasing, although the physical interpretation of this increase is not as obvious.

In the more generic region in the parameter space, $\Delta$ is not monotonic with increasing $m_{\chi}/m_{\phi}$. For example, $\Delta$ goes through a zero as $m_{\chi}/m_{\phi}$ is increased if $\alpha_{3}/\alpha_{4}\gtrsim \mathcal{O}(1)$.
Since $\Delta$ generically diverges as $m_{\chi}\rightarrow m_{\phi}$ and decreases with increasing $m_{\chi}$ for small $m_{\chi}/m_{\phi}$, there can be two zeroes if $\Delta>0$ when $m_{\chi}/m_{\phi}=0$ and $\alpha_{3}/\alpha_{4}\gtrsim \mathcal{O}(1).$ 

\section{\label{sec:Numerical-comparison}Numerical examples}

In this section, we employ the analytic results of \sref{subsec:Sample-analytic-amplitudes} to study GPP and quantum interference for two specific models of inflation: the Quadratic Potential model from \eref{eq:QuadraticModelDef} and the Hilltop Potential model from \eref{eq:HilltopModelDef}.  We evaluate the absolute value of the Bogoliubov coefficients $|\beta_k|$ using the analytic expressions for $\beta_k$ in \erefsss{eq:Bkn}{eq:betak_to_Ak_Phik}{eq:Ak}{eq:Phik}.  We consider a range of dimensionless comoving wavenumbers $k \in (10^{-2}, 10^2)$, where we've set $a_e H_e = 1$ such that the modes with $k = 1$ leave the horizon at the end of inflation.  

\begin{figure}
\includegraphics[width=0.90\textwidth]{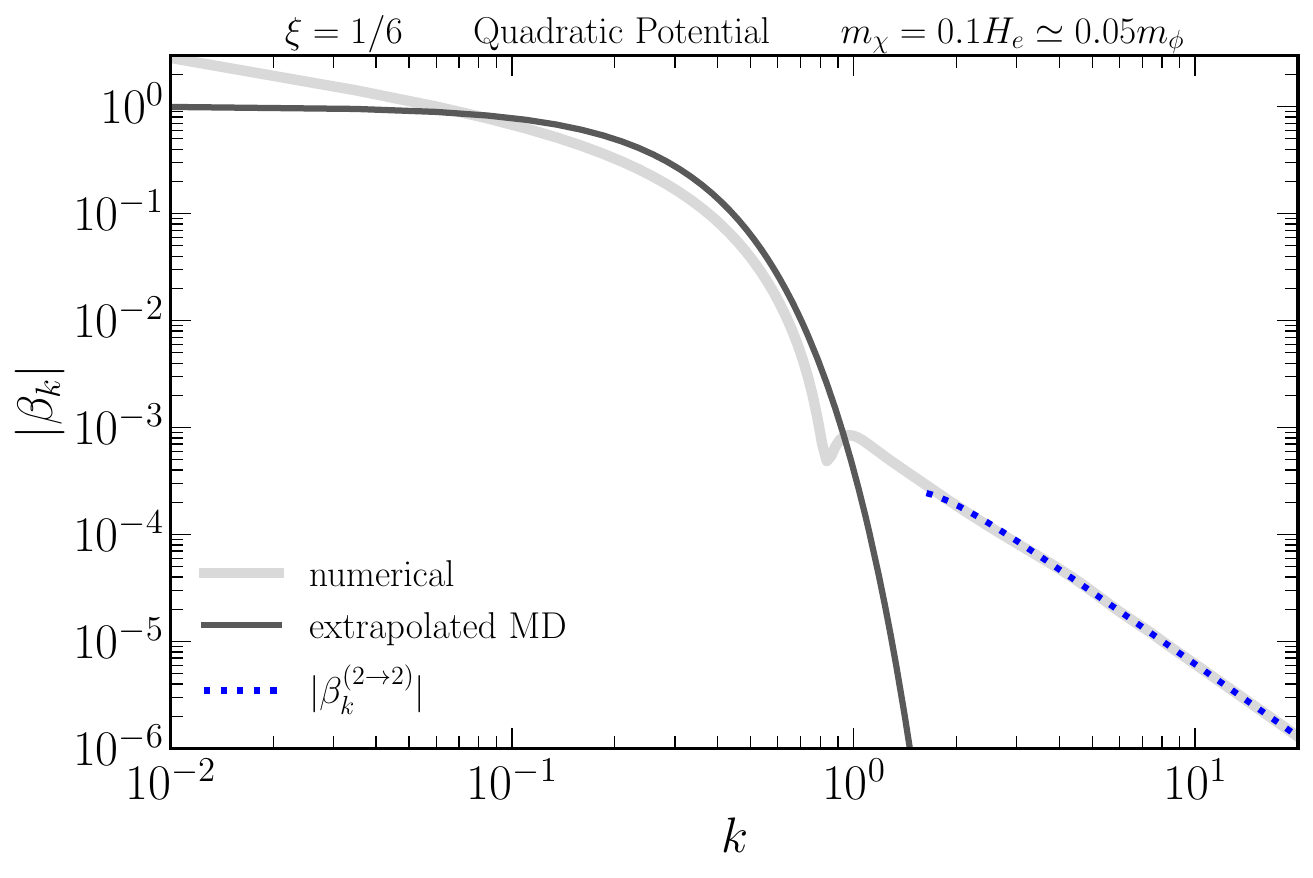} \\
\includegraphics[width=0.90\textwidth]{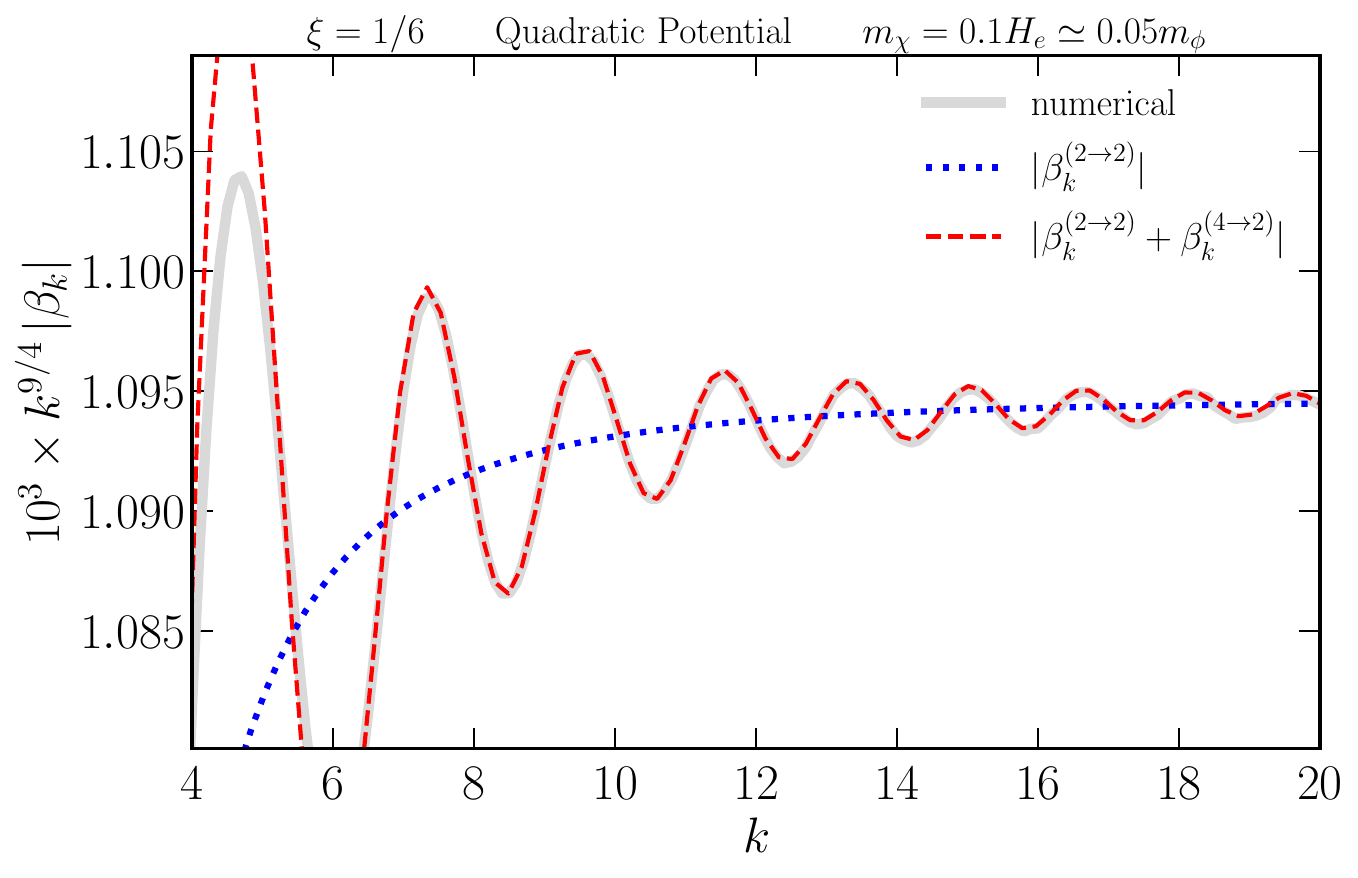} 
\caption{\label{fig:highlight-plot-quadratic}The Bogoliubov coefficient $|\beta_k|$ as a function of comoving wavenumber $k$ (with $a_e H_e = 1$) in the Quadratic Potential model.  We assume a conformally-coupled ($\xi=1/6$) scalar spectator with mass $m_\chi$ that experiences GPP due to an expanding spacetime background driven by an inflaton field $\phi$ on a quadratic potential with mass $m_\phi$.  \textit{Top:}  We calculate $|\beta_k|$ using the analytic results derived in this work (blue-dotted) and using direct numerical integration of the mode equations (gray).  Note that $|\beta_k|$ scales as $k^{-9/4}$ at large $k$.  As a comparative contrast to this power law behavior in $k$, the black curve shows an approximate expression for  $|\beta_k|$ for GPP in a matter dominated (MD) universe, extrapolated to lower $k$ values (beyond the range of strict validity) for visual completeness of the exponential behavior.     \textit{Bottom:}  The Bogoliubov coefficient exhibits an oscillatory feature in the large $k$ power-law tail of the spectrum, which is explained in this work as a result of quantum interference between $\beta_k^{(2\to2)}$ and $\beta_k^{(4\to2)}$. }
\end{figure}
 
\begin{figure}
\includegraphics[width=0.875\textwidth]{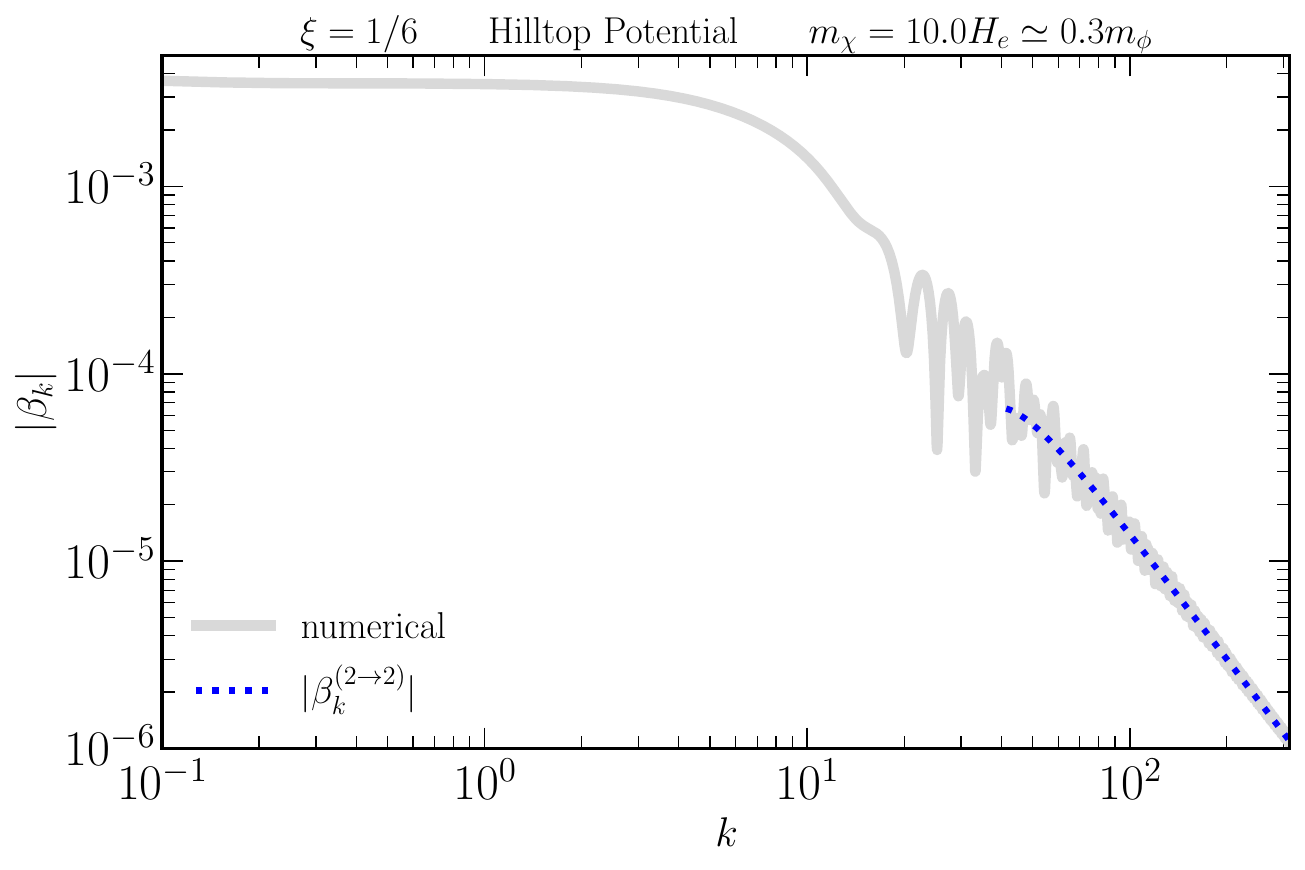}\\
\includegraphics[width=0.875\textwidth]{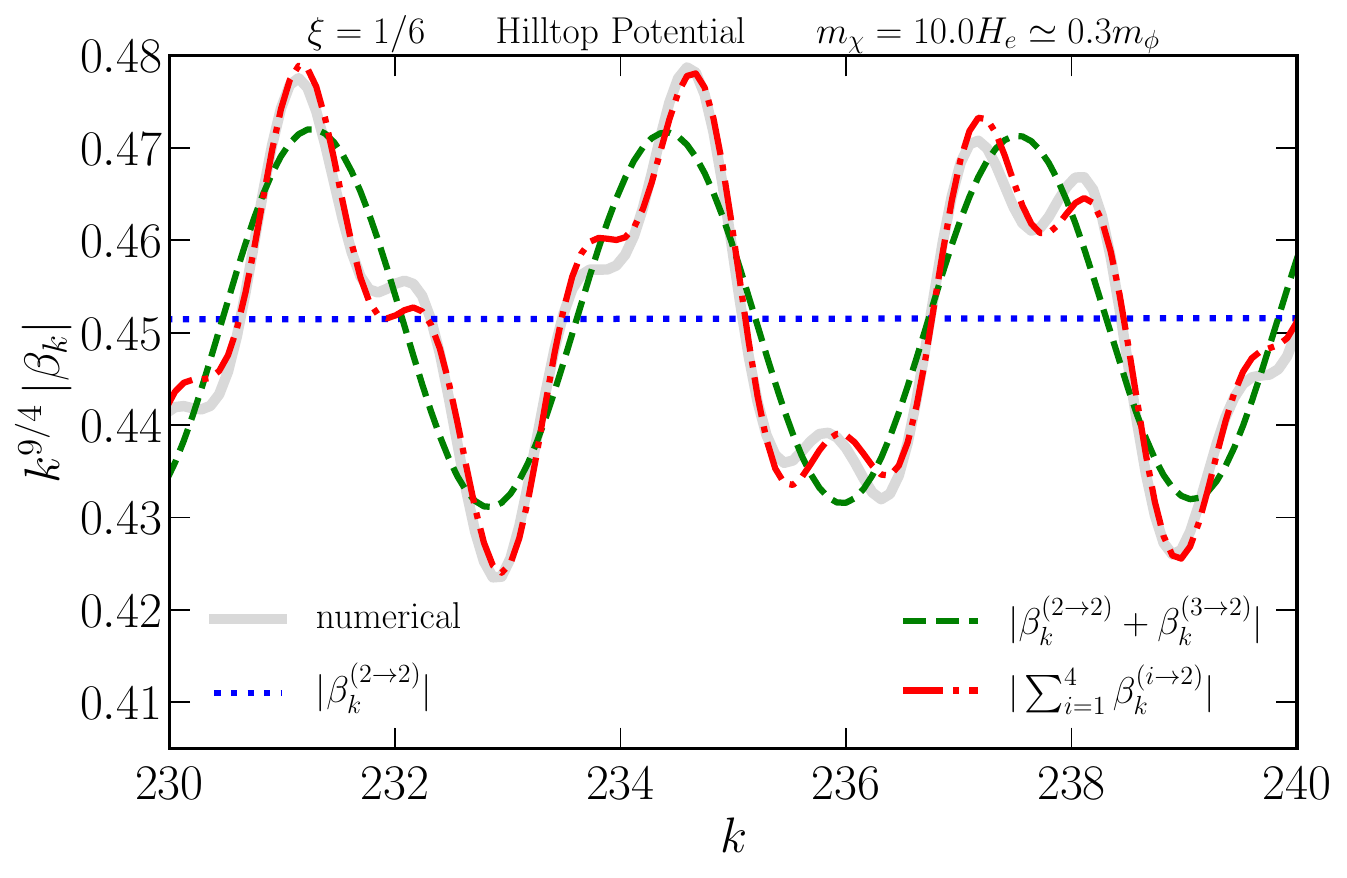}
\caption{\label{fig:highlight-plot-hilltop}Same as \fref{fig:highlight-plot-quadratic} but for the Hilltop Potential model and a different value of $m_\chi$.  Note the more pronounced and irregular oscillatory behavior, which is explained in this work as result of interference among four components: $\beta_k^{(1\to2)}$, $\beta_k^{(2\to2)}$, $\beta_k^{(3\to2)}$, and $\beta_k^{(4\to2)}$.  }
\end{figure}
 
Our results for the Quadratic Potential model are presented in \fref{fig:highlight-plot-quadratic}.  The blue-dotted curve corresponds to our leading-order analytic calculation $|\beta_k^{(2\to2)}|$ on both the upper and lower panels, while the red-dashed curve on the lower panel includes the first sub-leading correction  $|\beta_k^{(2\to2)}+\beta_k^{(4\to2)}|$.  Since the quadratic potential has a $\mathbb{Z}_2$ symmetry, $\phi \to - \phi$, the $n\phi$ processes with odd $n$ have vanishing amplitudes: e.g. $\beta_k^{(1\to2)} = 0$ and $\beta_k^{(3\to2)} = 0$.  The spectrum at large $k$ is approximately a power law $|\beta_k| \approx |\beta_k^{(2\to2)}| \propto k^{-9/4}$, but closer inspection reveals a sub-leading component that oscillates as $k$ is varied.  These oscillations are explained in this work as an interference effect. Using \eref{eq:leadingPhi}, the oscillation period $\Delta k$ is controlled by the variation in the phase with respect to $k$, and can be explicitly written as
\begin{equation}
\Delta k=\frac{2\pi}{|\partial_{k}(\Phi_{k}^{(n_{1}\rightarrow2)}-\Phi_{k}^{(n_{2}\rightarrow2)})|}\approx\frac{\frac{\pi}{2}\underline{a}m_{\phi}}{|n_{1}^{-1/2}-n_{2}^{-1/2}|}\sqrt{\frac{\underline{a}m_{\phi}}{2k}}
\end{equation}
for any $n_1$ and $n_2$. For the Quadratic model used in \fref{fig:highlight-plot-quadratic}, this evaluates to $\Delta k \simeq 2 a_e H_e$ for $n_1=2$ and $n_2=4$. For comparison, the gray curve shows the result of calculating $|\beta_k|$ by direct numerical integration of the mode equations.  The analytic results derived here agree very well with the numerical integration at large $k$ in both the average power-law behavior and the oscillatory features.  This agreement can be viewed as a validation of our analytic approximations.  The exponentially dropping black curve\footnote{The formula for this curve is
 $|\beta_k|=\exp{\left( - 2.47  (k/(a_e H_e))^{3/2} (H_e/m_\chi)^{1/2} \right)}$  valid for $k \gg a_e H_e$, and it corresponds to an approximate $|\beta_k|$ of GPP in a matter dominated universe.  It is 
 easily computable by several methods (e.g. \cite{Hashiba:2022bzi,Chung:1998bt}). A related formula is given explicitly in \cite{Ema:2018ucl}.} by contrast highlights the power-law behavior $|\beta_k| \propto k^{-9/4}$ coming from the oscillating inflaton field that drives corresponding oscillations in the scale factor.  

For the Hilltop Potential model, our results appear in \fref{fig:highlight-plot-hilltop}.  Once again, the leading power-law behavior at large $k$ is $|\beta_k| \approx |\beta_k^{(2\to2)}| \propto k^{-9/4}$ as seen from both the direct numerical integration (gray-solid) and our analytic approximation (blue-dotted).  The sub-leading oscillatory components (green-dashed and red-dot-dashed) have a richer structure in this model, which is evident by comparing the lower panels of figures~\ref{fig:highlight-plot-quadratic}~and~\ref{fig:highlight-plot-hilltop}.  This behavior can be understood as follows:  for the Hilltop Potential model the components $\beta_k^{(n\to2)}$ have similar amplitudes with increasing $n$, leading to a pronounced interference pattern, whereas the amplitudes decrease more rapidly in the Quadratic Potential model, and the interference is dominated by just the first two terms.  Moreover, since the Hilltop Potential model does not have a $\mathbb{Z}_2$ symmetry at the minimum of the inflaton's potential, the processes with an odd number of inflatons -- $\phi \to 2\chi$, $3\phi \to 2\chi$, and so on -- are not forbidden.  It turns out that $\beta_k^{(1\to2)}$ amplitude is numerically less important than that of $\beta_k^{(3\to2)}$ for the interference partly owing to the suppression of $(\kappa_3/\kappa_1)^{15/4}\ll1$ (see \eref{eq:epskappadef}).   By including up to the sub-sub-leading order in our analytic calculations, $|\beta_k^{(1\to2)} + \beta_k^{(2\to2)} + \beta_k^{(3\to2)} + \beta_k^{(4\to2)}|$, we obtain the red-dot-dashed curve that matches the result of direct numerical integration (gray-solid) very well at large $k$.    

Here we note the limits of applicability of our analytic results, using $\beta_k^{(\mathrm{2\to2})}$ as an example.   From the form of $\mathcal{A}_k^{(2\to2)}$, we see that for $m_\phi\gg m_\chi$ (which is the case for both figures) the next-order corrections to $\mathcal{A}^{(2\to2)}_k$ are approximately $x_1/1024\kappa_2^3$.  For the Quadratic Model $x_1/1024\simeq-1$, so the magnitude of the correction is approximately $\kappa_2^{-3}$.  For the Hilltop Model $x_1/1024\simeq-800$, and the magnitude of the correction is approximately  $800\kappa_2^{-3}$.  An upper limit on the magnitude of the correction results in a lower limit on $\kappa_2$, which, in turn results, in an $m_\phi/H_e$-dependent lower limit on $k$ (see \eref{eq:epskappadef}).  For the figures we have assumed that the next-order corrections to $\beta_k^{(\mathrm{2\to2})}$ are no more than $30\%$ (since the lower limit on $k$ only depends of the third-root of the correction limit, the result is relatively insensitive to the choice of $30\%$).   From the figures it is clear that the $k^{-9/4}$ behavior extends to $k$ somewhat lower than the cutoff in the convergence of our expansion.

\section{\label{sec:A-Heuristic-derivation}A heuristic derivation}

Our aim in this section is to describe semi-quantitatively the Bogoliubov computation of the resonance-induced GPP in terms of an approximate S-matrix perspective by showing how the Boltzmann equation would need to be modified to capture the interference effects. Here, we will focus on the interference of $2\rightarrow2$ and $3\rightarrow2$ scattering as this is often the most interesting case, with other generalizations being straightforward.

Consider an incoherent gas of $N\sim\rho_e V_3 / m_\phi$ number of $\phi$ particles, where $\rho_e\sim M_P^2 H_e^2$ is the energy density and $V_{3}\sim H_e^{-3}$ is the 3-volume of a large box approximating the causal Hubble patch. Usually, one first decoheres this large $N$ state system into an ensemble of $2\phi\rightarrow2\chi$ and $3\phi\rightarrow2\chi$, and then considers each process statistically independent.  In this case, the macroscopic particle production of $\chi$ is described by a semiclassical 1-particle $\chi$ distribution obtained from integrating the collision term as 
\begin{align}
& \int \partial_t f_{\chi}(k,t)dt \sim V_{3}^{3}\int\frac{d^{3}k_{2}}{(2\pi)^{3}}\frac{d^{3}p_{1}}{(2\pi)^{3}}\frac{d^{3}p_{2}}{(2\pi)^{3}}\mathcal{S}_{2}(p_{1},p_{2}) 
 \left|\langle\chi_{k}\chi_{k_{2}},t_{f}|U(t_{f},t_{e})|\phi_{p_{1}}\phi_{p_{2}},t_{e}\rangle\right|^{2} \nonumber \\
 & + V_{3}^{4}\int\frac{d^{3}k_{2}}{(2\pi)^{3}}\frac{d^{3}p_{1}}{(2\pi)^{3}}\frac{d^{3}p_{2}}{(2\pi)^{3}}\frac{d^{3}p_{3}}{(2\pi)^{3}}\mathcal{S}_{3}(p_{1},p_{2},p_{3})\left|\langle\chi_{k}\chi_{k_{2}},t_{f}|U(t_{f},t_{e})|\phi_{p_{1}}\phi_{p_{2}}\phi_{p_{3}},t_{e}\rangle\right|^{2}\label{eq:usual}
\end{align}
where $U(t_{2},t_{1})$ is the time-evolution operator from time $t_{1}$ to $t_{2}$, and $\mathcal{S}_{n}$ factors are $\phi$ initial-state dependent weighting factors (generalization of Bose-Einstein distribution), eventually leading to the cross section picture of the usual Boltzmann equations as shown explicitly in \aref{sec:obtainingboltzmann}. This treats ``typical'' 2-body scatterings and 3-body scatterings to be additive \emph{incoherently}. However, this type of computation neglects the nontrivial interference that can occur from Schr\"{o}dinger time evolution phases between different scatterings.

\begin{figure}
  \includegraphics[width=0.95\textwidth]{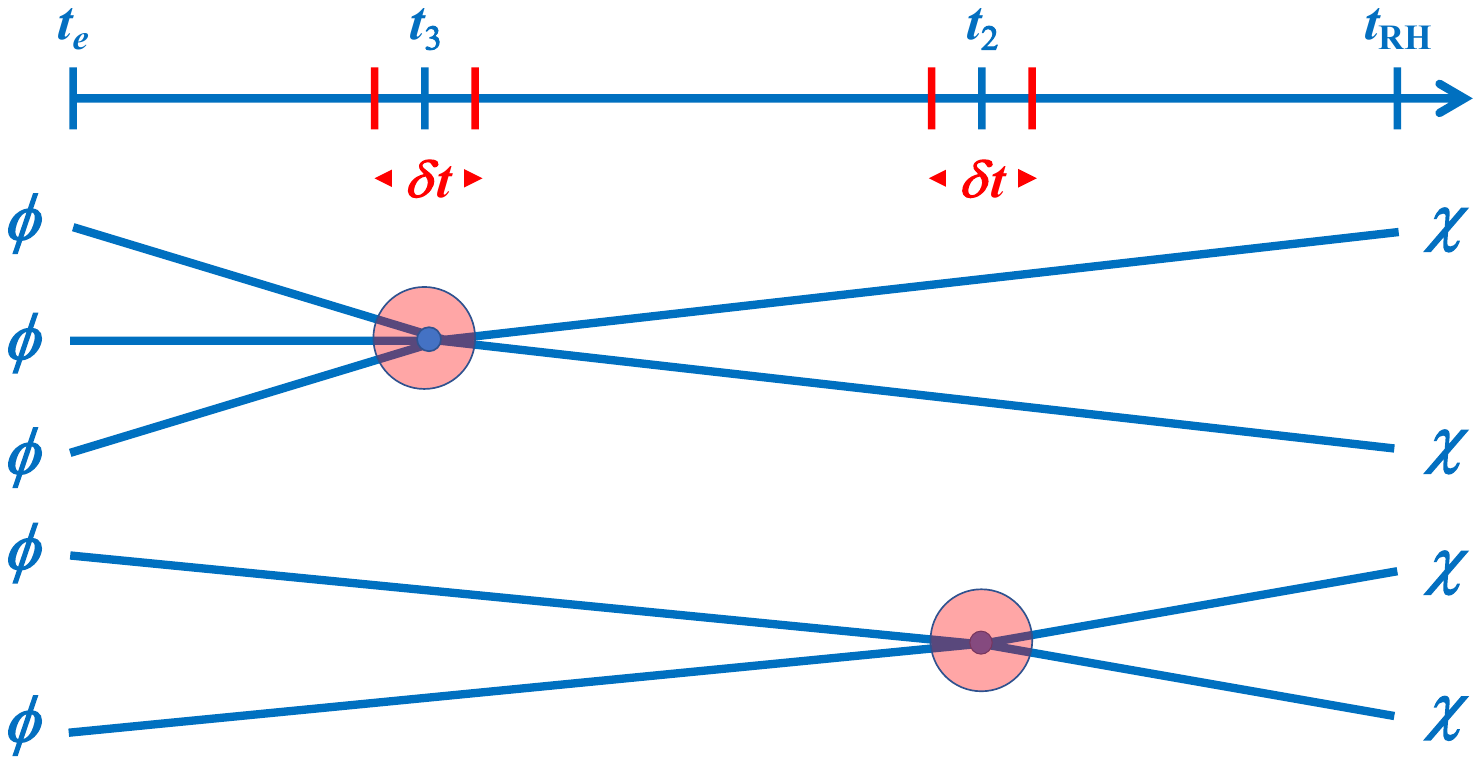}
  \caption{\label{fig:Schroedinger-propagator-phase} The Schr\"{o}dinger propagator phase difference between $3\phi\rightarrow 2\chi$ and $2\phi\rightarrow 2\chi$ scatterings leads to interference. The disk region of diameter $\delta t$ represents the usual collision region of Boltzmann equation, which is typically treated with an S-matrix taking the formal limit $\delta t\rightarrow\infty$. The observable interference phase of $\chi$ is the Schr\"{o}dinger-picture free-particle propagator between $t_{3}$ and $t_{2}$. The $\phi$-interference phase can contain $t_{e}$ information as 3$\phi$ propagation phase from $t_{e}$ to $t_{3}$ does not cancel the $2\phi$ propagation conjugate phase from $t_{e}$ to $t_{3}$.}
 \end{figure}

Hence, we arrive at the main idea. The scattering perspective that we will construct below will simply replace the nonadiabatic period during which the $2\chi$ particle frequencies are \emph{in resonance} with an approximate S-matrix scattering description. The different scattering events are diagrams (e.g., see \fref{fig:Schroedinger-propagator-phase}) that interfere because of the coherence of the waves entering the interaction region approximated by an S-matrix. This will allow us to compute the interference phase using the wave free-propagation phase. Thus, before we describe the scattering, let's divide the time period $[t_{e},t_{f}]$ into 3 regions:
\begin{equation*}
  \mbox{region 1:} \ [t_{e},t_{3}) \ , \qquad \mbox{region 2:} \ (t_{3},t_{2}) \ , \qquad  \mbox{region 3:} \ (t_{2},t_{f}] \ ,
 \end{equation*}
where $t_{n}$ is the time at which $n\phi\rightarrow2\chi$ resonance occurs, i.e., $2\sqrt{\frac{k^{2}}{a^{2}(t_{n})}+m_{\chi}^{2}}\approx nm_{\phi}$, which is the analog of the time $\bar{t}_k^{(n)}$ that satisfies the stationary-phase condition from \sref{sec:Bogoliubov-computation}.

From a scattering perspective, we work in the Schr\"{o}dinger picture with metric inhomogeneities in time, with $\phi$ treated as a quantum field, and the interaction Hamiltonian coming from the metric fluctuation coupling to the $\phi$ energy-momentum tensor. To describe the spatially homogeneous classical inflaton field, imagine setting up a normalized coherent state $|\phi)\sim$ (a state containing a macroscopic number of particles)\footnote{Here we follow the covention of \rref{Itzykson:1980rh} denoting a normalized coherent state as ``$|...)$''.} at time $t_{e}$ such that
\begin{equation}
 (\phi|U(t_{e},t)\,\phi(x,t)\, U(t,t_{e})|\phi)=\phi_{\mathrm{EOM}}(t) \label{eq:coherentstate_def}
\end{equation}
for $t>t_e$, where $\phi_{\mathrm{EOM}}(t)$ is the solution to \eref{eq:inflaton_EOM}. Note that the quantum phase of $U(t,t_{e})$ has turned into the classical phases embedded in $\phi_{\mathrm{EOM}}(t)$, approximated as $n m_\phi t$ for integers $n$. This is one source of the interference phase as we will see below.

We will assume for the semi-quantitative discussion that this coherent state can be generalized straightforwardly to the effective FLRW background of an expanding box: $ds^{2}=dt^{2}-a_{\mathrm{slow}}^{2}(t)|d\vec{x}|^{2}$. Instead of treating $H_{\mathrm{slow}}$ as a Minkowski graviton effect, the background is treated as an expanding box even in the scattering picture because a purely Minkowski treatment is inefficient in explaining $k/a_{\mathrm{slow}}$ dilution. 

The normalized coherent state $|\phi)$ of \eref{eq:coherentstate_def} can be decomposed as a superposition of normalized wave packet states over $r$ numbers of $\phi$ particles, written as
\citep{Itzykson:1980rh}
\begin{align}
 |r,\{q_{i}\},t_{e}\rangle_{\phi} & =\frac{\int[dk_{1}]...[dk_{r}]F_{(r)}^{(\{q_{i}\})}(k_{1},...,k_{r})a_{\vec{k}_{1}}^{\dagger}...a_{\vec{k}_{r}}^{\dagger}|0,t_{e}\rangle_{\phi}}{\sqrt{r!\int[dk_{1}]...[dk_{n}]|F_{(r)}^{(\{q_{i}\})}(k_{1},...,k_{r})|^{2}}} \label{eq:wavepacket} \ ,
\end{align}
where $F_{(r)}^{(\{q_{i}\})}$ controls the $\phi$ particle wave packets with central momenta $\{q_{i}\}\equiv\{q_{1},...,q_{r}\}$. 
For illustration, suppose the initial state in the notation of \eref{eq:wavepacket} is decohered into clusters of $2$ and $3$-particle states described by a density matrix $\rho=\sum_{\psi}\mathcal{S}(\psi)|\psi\rangle\langle\psi|$, where 
\begin{align}
|\psi\rangle & \approx |0,t_{e}\rangle_{\chi}\otimes|\Psi\rangle_{\phi}\otimes|0,t_{e}\rangle_{\delta g_{\mu\nu}} \ ,
\end{align}
\begin{align}
|\Psi\rangle_{\phi} & \equiv \frac{ \zeta_{2}|2,\{p_{i}\},t_{e}\rangle_{\phi}+\zeta_{3}|3,\{p_{i}\},t_{e}\rangle_{\phi}}{\sqrt{|\zeta_{2}|^{2}+|\zeta_{3}|^{2}}}\label{eq:illustrativepartition} \ ,
\end{align}
and $\mathcal{S}$ partitions the macroscopic $N$-inflaton state into an ensemble of coherent superpositions of 2-particle and 3-particle states. The wave packet function $F_{r}$ appearing in \eref{eq:wavepacket} is assumed to be peaked at close to zero spatial momentum since the inflatons are assumed to be cold.  The amplitudes $\zeta_{2}$ and $\zeta_{3}$ control the mixing of 2- and 3-particle inflaton states.
The $\phi$ state of \eref{eq:illustrativepartition} can be intuitively considered a ``classical'' coherence because it represents a macroscopic state,\footnote{The sum over $\mathcal{S}$ is a macroscopic number, similar to $V_{3}^{3}\int\frac{d^{3}k_{2}}{(2\pi)^{3}}\frac{d^{3}p_{1}}{(2\pi)^{3}}\frac{d^{3}p_{2}}{(2\pi)^{3}}\mathcal{S}_{2}(p_{1},p_{2})$ in \eref{eq:usual}.} and the Bogoliubov vacuum does not contain the quantum data for $\phi$ in $\zeta_{n}$. However, this ``classical'' coherence itself is really part of the quantum coherence associated with the Schr\"{o}dinger time evolution operator just as in photon time-phase coherence in lasers. 

With the illustrative partition of \eref{eq:illustrativepartition}, the analog of \eref{eq:usual} becomes
\begin{equation}
  \int\partial_{t}f_{\chi}(k,t)dt=\frac{2N_{\mathrm{partition}}[\mathcal{S}]}{|\zeta_{2}|^{2}+|\zeta_{3}|^{2}}\left|\langle\chi_{k}\chi_{k_{2}},t_{f}|U(t_{f},t_{e})\left(\zeta_{2}|2,\{p_{i}\},t_{e}\rangle+\zeta_{3}|3,\{p_{i}\},t_{e}\rangle_\phi\right)\right|^{2} \label{eq:boltzmann_analog}
\end{equation}
where $N_{\mathrm{partition}}[\mathcal{S}]$ is a multiplicity factor associated with the partition achieved through the density matrix probability factor $\mathcal{S}$. Because of the resonant behavior, we know
\begin{align}
  \langle\chi_{k}\chi_{k_{2}},t_{f}|U(t_{f},t_{e})|n,\{p_{i}\},t_{e}\rangle\approx e^{-2i\int_{t_{n}^{+}}^{t_{f}}E_{k}(t)dt}\mathcal{A}_{n}\,e^{-ni\int_{t_{e}}^{t_{n}^{-}}m_{\phi}dt} \ ,
\end{align}
\begin{align}
  \mathcal{A}_{n}\equiv\langle\chi_{k}\chi_{k_{2}},t_{n}^{+}|U(t_{n}^{+},t_{n}^{-})|n,\{p_{i}\},t_{n}^{-}\rangle \ ,
\end{align}
where $t_{n}^{\pm}\equiv t_{n} \pm \delta t /2$, with $\delta t$ as the interaction time, i.e., the time scale of a Boltzmann collision term, which by construction is supposed to be much smaller than the free-streaming time scale. However, $\delta t$ is viewed in the S-matrix picture as an asymptotically long time scale,  as one formally takes $\delta t\rightarrow\infty$ to take advantage of the properties associated with meromorphic matrix elements.\footnote{Note that $\mathcal{A}_k^{(n\to2)}$ is typically proportional to the scattering amplitude and $\mathcal{A}_n$ at tree level order.}  This is the usual requirement of the validity of the Boltzmann treatment.  Hence, the squared amplitude in the modified Boltzmann collision analog of \eref{eq:boltzmann_analog} becomes
\begin{align}
  \left|\zeta_{2}\mathcal{A}_{2}\right|^{2}+\left|\zeta_{3}\mathcal{A}_{3}\right|^{2}+2\Re[e^{-2i\int_{t_{3}^{+}}^{t_{2}^{+}}E_{k}(t)dt+2i\int_{t_{e}}^{t_{2}^{-}}m_{\phi}dt-3i\int_{t_{e}}^{t_{3}^{-}}m_{\phi}dt}\zeta_{2}^{*}\zeta_{3}\mathcal{A}_{2}^{*}\mathcal{A}_{3}] \ , \label{eq:interfering-phase}
\end{align}
where one notes in \eref{eq:interfering-phase} that the cross-term induced coefficient as part of the interference phase. More generically, the interference phase between $n_{1}\phi\rightarrow2\chi$ and $n_{2}\phi\rightarrow2\chi$ is
\begin{equation}
 -2\int_{t_{n_{1}}}^{t_{n_{2}}}E_{k}(t)dt+n_{2}\int_{t_{e}}^{t_{n_{2}}}m_{\phi}dt-n_{1}\int_{t_{e}}^{t_{n_{1}}}m_{\phi}dt \label{eq:interferencephase}
\end{equation}
in the limit that $\delta t\ll |t_{n_{2}}-t_{n_{1}}|$. The phase of the Schr\"{o}dinger-propagator independent quantity $\zeta_{n_1}^{*}\zeta_{n_2}$ is apparently independent of $n$ in the case of our particular $|\beta_{k}|^{2}$ computation.  

Comparing with \eref{eq:interference-phase}, we see that the $\hypgeo{2}{1}$-proportional pieces in $\Phi_{k,\mathrm{leading}}^{(2\rightarrow2)}-\Phi_{k,\mathrm{leading}}^{(3\rightarrow2)}$ matches $-2\int_{t_{3}}^{t_{2}}E_{k}(t)dt$, where the hypergeometric function arises from integrals of the form
\begin{equation}
  \int_{t_{3}}^{t_{2}}E_{k}(t)dt \approx\int_{\kappa_{3}^{-3/2}}^{\kappa_{2}^{-3/2}}\frac{d\lambda}{-\frac{3}{2}m_{\phi}\lambda^{2}}\sqrt{\frac{k^{2}}{\underline{a}^{2}}\lambda^{4/3}+m_{\chi}^{2}} \ , \label{eq:energyintegral}
\end{equation}
which used the relationship $a \approx \underline{a}\lambda^{-2/3}$ and the definition of $\dot{\lambda}$ in \eref{eq:lambda_dot}. The hypergeometric function term by itself has a divergent piece as $m_{\chi}\rightarrow0$, which is obviously spurious since the left hand side of \eref{eq:energyintegral} is convergent for finite $t_{n}$. Similarly, the remaining terms of \eref{eq:interference-phase} can be identified with the inflaton phase:
\begin{equation}
(2-3)\left(\underline{\Xi} - \frac{2m_{\phi}}{3\underline{H}}\right)+\frac{2}{3}\left(2\kappa_{2}^{3/2}-3\kappa_{3}^{3/2}\right)\longleftrightarrow2\int_{t_{e}}^{t_{2}}m_{\phi}dt-3\int_{t_{e}}^{t_{3}}m_{\phi}dt \ ,
\end{equation}
which also matches the interpretation of $\underline{\Xi}$ being the phase offset that depends on the properties of the inflaton at the end of the quasi-dS era at time $t_{e}$.

\section{\label{sec:conclusion}Conclusions}

In this article we report on our study of quantum interference in the phenomenon of gravitational particle production.  
Our main results appear in \sref{subsec:Sample-analytic-amplitudes}.  
We have derived analytic expressions for the Bogoliubov coefficients $\beta_k$ describing the gravitational production of conformally-coupled, massive scalar particles during the inflaton's coherent oscillations after inflation.  
By employing a novel perturbation technique (relying on a nonlinear field redefinition) and a stationary phase calculation, we have expressed $\beta_k$ as a sum over resonant contributions $\beta_k^{(n\to2)}$.  
Oscillatory features in the spectrum $|\beta_k|^2$ are understood to result from an interference among the resonant contributions, e.g. $|\beta_k^{(2\to2)} + \beta_k^{(3\to2)}|^2 \neq |\beta_k^{(2\to2)}|^2 + |\beta_k^{(3\to2)}|^2$; see also \eref{eq:interference-phase} for details.  
These analytic results are in excellent agreement with a direct numerical integration of the mode equations; as shown in \sref{sec:Numerical-comparison}, the agreement is within a few percent in certain kinematic regions.  
Our work explains much of the previously unexplained ``noise'' in numerically-computed spectra, seen for example in refs.~\citep{Giudice:1999am,Ema:2018ucl,Kolb:2021xfn}.  
As we discuss in \sref{sec:A-Heuristic-derivation}, the resonant contributions $\beta_k^{(n\to2)}$ are related to gravity-mediated inflaton scattering amplitudes $n\phi \to 2\chi$ corresponding to $n$ inflaton particles with mass $m_\phi$ at rest annihilating to $2$ scalar particles with mass $m_\chi < n m_\phi / 2$.  
This work also elucidates the quantum nature of gravitational particle production induced by classical inflaton coherent dynamics.  

As noted in \sref{sec:A-Heuristic-derivation}, the interference phase can be understood as arising from the free propagator phases of the external legs of the scattering process. This means that the phases are dependent on the kinematics of the inflaton and the $\chi$ particles, as well as the scattering times of say $n_{1}\phi\rightarrow2\chi$ and $n_{2}\phi\rightarrow2\chi$ processes. Unlike the usual scattering situations where $n_{1}\phi\rightarrow2\chi$ and $n_{2}\phi\rightarrow2\chi$ are incoherent, the coherent oscillation nature of the initial inflaton state allows for the scattering amplitudes to interfere. This interference is efficiently captured using the Bogoliubov transformation formalism. 

The modulations of the $\chi$-particle momentum spectrum shown in figures~\ref{fig:highlight-plot-quadratic}~and~\ref{fig:highlight-plot-hilltop} in principle can be probed by kinematic-dependent subsequent scattering dynamics of $\chi$ particles. For example, if interesting motivated scenarios exist for $\chi$ particle scattering resonances with judicious energy spacing, the interference pattern of $\chi$ energies may lead to enhanced production of final states compared to situations without this interference pattern in the $\chi$ particle spectrum. Investigations into possible applications will be left to future work.

\begin{acknowledgments}
E.B.\ was supported in part by the generosity of the Ray MacDonald fund during this work.  E.W.K. was supported in part by the US Department of Energy contract DE-FG02-13ER41958.  A.J.L. was support in part by the National Science Foundation under award number PHY-2114024.
\end{acknowledgments}

\appendix

\section{\label{sec:radiaangular}Background field evolution in novel polar coordinates}

The evolution of the inflaton field $\phi$ is usually described by the second-order equation
\begin{equation}
  \ddot{\phi}+3\dot{\phi}\sqrt{\frac{\dot{\phi}^2+2V(\phi)}{6M_P^2}} + V'(\phi) = 0 \ , \label{eq:inflaton_EOM}
\end{equation}
which is often referred to as the inflaton equation of motion. For our purposes we wish to exchange the second-order differentiation to a set of two first-order equations at the cost of introducing another dependent variable.  There is freedom in choosing the two variables; our choice is the Hubble rate $H$ and phase $\Xi$ (both reals), defined in terms of $\phi$ and $\dot{\phi}$ as
\begin{equation}
 \boxed{\sqrt{6}M_P H e^{i\Xi} \equiv  s_\phi\sqrt{2V(\phi)}-i\dot{\phi} \ , } \label{eq:HXi_def}
\end{equation}
where $s_\phi \equiv \mathrm{sign}(\phi-v)$ and $v$ is the field value where the potential is minimized. As explained in \rref{preplamtheta}, the change of variables from $\{\phi,\dot{\phi}\} \to \{H,\Xi\}$ is analogous to switching from Cartesian to polar coordinates in phase space, with $H$ representing the radial coordinate and $\Xi$ the angular coordinate.

From \erefs{eq:inflaton_EOM}{eq:HXi_def}, the equations of motion (EOMs) for $H$ and $\Xi$ are
\begin{equation}
 \dot{H}=-3H^{2}\sin^{2}\Xi\ ,\qquad\dot{\Xi}=\frac{V'(\phi_{H,\Xi})}{\sqrt{6}M_{P}H\cos\Xi}-\frac{3}{2}H\sin2\Xi\ , \label{eq:EOMs_for_H_Xi}
\end{equation}
respectively, where $\phi_{H,\Xi}$ is the solution to
\begin{align}
 \sqrt{6} M_P H \cos\Xi= s_{\phi_{H,\Xi}} \sqrt{2 V(\phi_{H,\Xi})} \label{eq:Hcos} \ ,
\end{align}
which is simply the real part of \eref{eq:HXi_def}. As done in \rref{preplamtheta}, we simplify the presentation of our problem by using the change of variables
\begin{equation}
  \boxed{\phi \to v + \sqrt{6} M_P \phi \ , \qquad t \rightarrow t_e + m_\phi^{-1} t \ ,} \label{eq:dimensionless_def}
\end{equation}
which is equivalent to setting $\sqrt{6}M_P = m_\phi = 1$ and $v=t_e=0$. Using the expansion of the potential in \eref{eq:Vpert}, we write \eref{eq:Hcos} as
\begin{equation}
  H\cos\Xi = \phi_{H,\Xi} \sqrt{1 + 2 \alpha_3 \phi_{H,\Xi} + 2 \alpha_4 \phi_{H,\Xi} ^2 + \dots} \ ,
\end{equation}
and invert this to find 
\begin{align}
  \phi_{H,\Xi} = H\cos\Xi\left(1 - \alpha_3 H \cos\Xi + \left(\frac{5}{2}\alpha_3^2-\alpha_4\right)H^2\cos^2\Xi + \dots \right)  \ , \label{eq:phiminusphimin}
 \end{align}
which allows us to express the EOMs entirely in terms of $H$ and $\Xi$:
\begin{equation}
  \frac{dH^{-1}}{dt} = \frac{3}{2}(1-\cos2\Xi) \ , \qquad \frac{d\Xi}{dt} = 1-\frac{3}{2}H\sin2\Xi+\sum_{n=1}^{\infty}\nu_{n}H^{n}\cos^{n}\Xi \ , \label{eq:EOMs_for_H_Xi_dimensionless}
\end{equation}
for some constants $\nu_n$, with $\nu_1=2\alpha_3$, $\nu_2=-\frac{7}{2}\alpha_{3}^{2}+3\alpha_{4}$, and so on. We write the derivatives for $H^{-1}$ and $\Xi$ to highlight that both grow linearly with time if $H$ is small. This will be useful when deriving the constants $\underline{H}$ and $\underline{\Xi}$, which are associated with the boundary conditions at $t=+\infty$ for $H$ and $\Xi$, respectively.

\subsection{Defining boundary conditions in asymptotic far future}

This subsection introduces the constants $\underline{H}$, $\underline{\Xi}$, and $\underline{a}$ that our formulas for $\beta_k^{(n\to2)}$ ultimately depend on. These constants quantify the boundary conditions in the far future as $t\to+\infty$, and contain the same information as the initial conditions. This limit is necessary to integrate the equations of motion starting at $\lambda=0$ for our asymptotic expansions. We start by noting that
\begin{equation}
  \frac{d}{dt} \left(H^{-1} - \frac{3}{2}t+\frac{3}{4}\sin2\Xi\right) \sim \frac{d}{dt}\left( \Xi - t \right) \sim \frac{d}{dt}\log(H^{2/3}a) \sim \mathcal{O}(H) \ \label{eq:late_time_behavior}
\end{equation}
is a consequence of the EOMs in \eref{eq:EOMs_for_H_Xi_dimensionless}. We see that the quantities in parenthesis approach constants as $H\to0$ at late times.\footnote{We acknowledge that $\log H$ behavior at late times is consistent with \eref{eq:late_time_behavior}. However the $\lambda,\theta$ asymptotic expansions show such divergences do not exist for the quantities in the parenthesis.} We define them as
\begin{align}
  \underline{H}^{-1} & \equiv \lim_{t\to \infty} H^{-1}(t)-\frac{3}{2}(t-t_e) + \frac{3}{4m_\phi}\sin2\Xi(t) \ , \label{eq:H_underline_def} \\
  \underline{\Xi} & \equiv \lim_{t\to\infty}\,\Xi(t) - m_\phi (t-t_e)  \ ,  \label{eq:Xi_underline_def} \\
  \underline{a} & \equiv \lim_{t\to \infty}\,a(t)\,(H(t)/m_\phi)^{2/3} \ , \label{eq:a_underline_def}
\end{align}
where we restored units by reversing \eref{eq:dimensionless_def}.

While these constants can be determined by numerical integration of the background field equations, they can also be estimated using an expansion in $H(t)/m_\phi$. Using dimensional reduction of \eref{eq:dimensionless_def}, we write our three dynamical variables as
\begin{align}
  H^{-1}(t) & =\underline{H}^{-1}+\frac{3}{2}(t-t_{e})-\frac{3}{4}\sin2\Xi(t)+\sum_{n=1}^{\infty}H^{n}(t)R_{n}\left(\Xi(t)\right) \ , \nonumber  \\
  \Xi(t) & =\underline{\Xi}+t-t_{e}+\sum_{n=1}^{\infty}H^{n}(t)X_{n}\left(\Xi(t)\right) \ , \nonumber \\
  a(t) & =\underline{a}H^{-2/3}(t)\exp\left(\frac{H}{2}\sin2\Xi+\sum_{n=1}^{\infty}H^{n+1}(t)A_{n}(\Xi(t))\right) \ ,
\end{align}
for some oscillatory functions $R_n(\Xi)$, $X_n(\Xi)$ and $A_n(\Xi)$,  which we obtain by solving \eref{eq:EOMs_for_H_Xi_dimensionless} along with $\frac{d}{dt}\log a = H$ at each order in $H$. The constants of integration are again determined by the condition of eliminating divergences to keep these functions bounded. Unlike in the case of the $\lambda,\theta$ expansion, all constants are completely determined such that the solution is unique with no scheme choices needed.

By simple subtraction, this solution immediately yields expressions for the boundary constants of \erefss{eq:H_underline_def}{eq:Xi_underline_def}{eq:a_underline_def} based on the values of $H$, $\Xi$ and $a$ at any target time $t$. If we choose this time to be $t_e$, then to first order in $H_e/m_\phi$ we have the estimates
\begin{align}
  \frac{m_{\phi}}{\underline{H}} & \approx \frac{m_{\phi}}{H_{e}}+\frac{3}{4}\sin2\Xi_{e}-\frac{H_{e}}{m_{\phi}}\left(\frac{15\alpha_{3}^{2}}{8}-\frac{3\alpha_{4}}{4}+\frac{9}{32}\cos4\Xi_{e}+\frac{3\alpha_{3}}{2}\sin\Xi_{e}+\frac{\alpha_{3}}{2}\sin3\Xi_{e}\right) \ , \nonumber \\
  \underline{\Xi} & \approx\Xi_{e}-\frac{H_{e}}{m_{\phi}}\left(\frac{9}{8}+\frac{5\alpha_{3}^{2}}{2}-\alpha_{4}+\frac{3}{4}\cos2\Xi_{e}+2\alpha_{3}\sin\Xi_{e}\right) \ , \nonumber  \\
  \frac{\underline{a}}{a_{e}} & \approx\frac{H_{e}^{2/3}}{m_{\phi}^{2/3}}\exp\left[-\frac{H_{e}}{2m_{\phi}}\sin2\Xi_e -\frac{H_{e}^{2}}{m_{\phi}^{2}}\left(-\frac{3}{8}-\frac{5\alpha_{3}^{2}}{8}+\frac{\alpha_{4}}{4}  \right. \right. \nonumber \\ &  \left. \left. \hspace*{72pt} -\frac{3}{8}\cos2\Xi_{e}-\frac{3}{32}\cos4\Xi_{e}-\alpha_{3}\sin\Xi_{e}-\frac{1}{3}\alpha_{3}\sin3\Xi_{e}\right)\right]  \ .
\end{align}

\subsection{\label{subsec:Xi_end}A special property at the end of inflation}

Here we note a special property about $\Xi_e$. The end of inflation is defined by
\begin{equation}
  \ddot{a}_e=0\quad\leftrightarrow\quad\rho_e+3p_e=0\quad\leftrightarrow\quad\dot{\phi}_e^2=V(\phi_e) \ ,
\end{equation}
where the arrows represent equivalence between all three statements. Using the third statement and \eref{eq:HXi_def}, we can show that
\begin{equation}
 \Xi_e = \arg\left[s_e(\sqrt{2}+i)\right] = \pm\frac{\pi}{2}-\arctan\sqrt{2}, \label{eq:Xi_end_geometric}
\end{equation}
where $s_e\equiv\mathrm{sign}(\phi_e - v)=\pm1$. This has the advantage of specifying the end of inflation in a closed-form and geometric manner using a single variable with no derivatives involved. Examples of both physically distinct solutions for $\Xi_e$ can be found in \tref{tab:Hbar_Xibar_abar_table}.

Using the geometric expression of $\Xi_e$ in \eref{eq:Xi_end_geometric}, we can express the approximations for the boundary condition constants from the previous subsection as
\begin{align}
 \frac{m_{\phi}}{\underline{H}} &	= \frac{m_{\phi}}{H_{e}}+\frac{1}{\sqrt{2}}+\frac{H_{e}}{m_{\phi}}\left(\frac{7}{32}-\frac{15\alpha_{3}^{2}}{8}+\frac{3\alpha_{4}}{4}-\frac{7s_e\alpha_{3}}{3\sqrt{3}}\right) \ , \label{eq:H_underline_analytic_estimate} \\
 \underline{\Xi} & =\arg\left[s_e(\sqrt{2}+i)\right]+\frac{H_{e}}{m_{\phi}}\left(-\frac{11}{8}-\frac{5\alpha_{3}^{2}}{2}+\alpha_{4}-\frac{2s_e\alpha_{3}}{\sqrt{3}}\right) \ , \label{eq:Xi_underline_analytic_estimate} \\
 \frac{\underline{a}}{a_{e}}	 & =\frac{H_{e}^{2/3}}{m_{\phi}^{2/3}}\exp\left[-\frac{\sqrt{2}H_{e}}{3m_{\phi}}+\frac{H_{e}^{2}}{m_{\phi}^{2}}\left(\frac{41}{96}+\frac{5\alpha_{3}^{2}}{8}-\frac{\alpha_{4}}{4}+\frac{14s_e\alpha_{3}}{9\sqrt{3}}\right)\right] \ , \label{eq:a_underline_analytic_estimate}
\end{align}
which only uses the initial conditions of $a_e$, $H_e$, and $s_e$, along with information about the potential up to the quartic interaction, i.e., $m_\phi$, $\alpha_3$, and $\alpha_4$. We expect $s_e$ to only appear with quantities such as $\alpha_3$ that break the reflection symmetry of the potential about its minimum, i.e., $V(v+\Delta\phi)=V(v-\Delta\phi)$, i.e., $\alpha_n=0$ for all odd $n$. If the potential is symmetric, then $\underline{H}$, $\underline{\Xi}-\Xi_{e}$, and $\underline{a}$ must be independent of the $s_e$ initial condition. In this case, the action $s_e\rightarrow-s_e$ causes both $\Xi_{e}$ and $\underline{\Xi}$ shift by the same factor of $\pi$ such that the difference is unaffected.

\section{\label{sec:Small--expansion}Summary of the perturbative asymptotic series formalism}

This section summarizes the formalism of \rref{preplamtheta}, in which the equations of motion are solved using asymptotic expansions of the variables as a function of ``slow-'' and ``fast-'' time variables $\lambda$ and $\theta$. As shown in \sref{sec:Bogoliubov-computation}, this allows computation of the Bogoliubov coefficients as an analytic expansion in powers of $k^{-3/2}$.

We begin by using \eref{eq:dimensionless_def} to scale and shift away various constants. It is convenient to solve the dynamics using the polar coordinates $H$ and $\Xi$ defined by \eref{eq:HXi_def}, as the former is used in GPP calculations. We use perturbative series in powers of $\lambda$ to write them as
\begin{align}
 H & = \lambda+\sum_{\ell=1}^{\infty}h_{\ell}(\theta)\lambda^{\ell+1} \label{eq:lam_expansion_of_H} \ , \\
 \Xi & =\theta+\sum_{\ell=1}^{\infty}\xi_{\ell}(\theta)\lambda^{\ell}\label{eq:lam_expansion_of_Xi} \ ,
\end{align}
where $h_\ell(\theta)$ and $\xi_\ell(\theta)$ are oscillatory functions that must remain bounded in magnitude to maintain the stability of the expansion. This requirement will determine most of the constants of integration associated with solving for these functions at each order in $\lambda$. As one can see from \eref{eq:lam_expansion_of_H}, the expansion in powers of $\lambda$ is justified if $H/m_{\phi}$ is small.

To solve the EOMs from \eref{eq:EOMs_for_H_Xi_dimensionless}, we must specify the time evolution of $\lambda$ and $\theta$. We define the derivatives of both variables as
\begin{align}
 \dot{\lambda} & = \beta_\lambda(\lambda) \equiv -\frac{3}{2}c_0\lambda^{2}\left(1+\cancel{c_1\lambda}+\sum_{j=2}^{\infty}c_j\lambda^j\right) \ , \label{eq:lambda_dot} \\
 \dot{\theta} & = \beta_\theta(\lambda) \equiv \omega_0\left(1+\cancel{\omega_1\lambda}+\sum_{j=2}^{\infty}\omega_j\lambda^j\right) \ , \label{eq:theta_dot}
\end{align}
where $c_j$ and $\omega_j$ are constant coefficients. While the EOMs determine $c_0=\omega_0=1$ and $c_1=\omega_1=0$, the coefficients for all $j\geq2$ remain unfixed parameters. It will be shown that the parameter
\begin{equation}
  \underline{h}_{1}\equiv\frac{1}{2\pi}\int_{0}^{2\pi}h_{1}(\theta)d\theta
 \end{equation}
also remains undetermined due to the time translation invariance of the EOMs, i.e., the freedom to choose the origin of the time coordinate. We call every such choice of $\underline{h}_1$ and $\{c_{j},\omega_{j}|j\geq2\}$ a renormalization scheme (RS) because of the analogy with the coupling flow equations. Of course, given that this degree of freedom is a diffeomorphism choice, we could have also called it a gauge choice.

We are free to choose $\underline{h}_1$ by the following argument. The derivatives in \erefs{eq:lambda_dot}{eq:theta_dot} are invariant under shifts in $t$ or $\theta$. Thus, if $\lambda(t)$ and $\theta(t)$ are solutions for a given set of coefficients $\left\{ c_{j},\omega_{j}\right\} $, then $\lambda(t-t_{s})$ and $\theta(t-t_{s})-\theta_{s}$ must also be solutions for any shifts $t_{s}$ and $\theta_{s}$. Using the typical Taylor series, this can be expressed as making the replacements
\begin{align}
 \lambda & \rightarrow\lambda+\frac{3}{2}t_{s}\lambda^{2}+\frac{9}{4}t_{s}^{2}\lambda^{3}+\mathcal{O}(\lambda^{4}) \ , \\
 \theta & \rightarrow\theta-\theta_{s}-t_{s}+\omega_{2}t_{s}\lambda^{2}+\mathcal{O}(\lambda^{3}) \ , 
\end{align}
with derivatives at $t$ determined by \erefs{eq:lambda_dot}{eq:theta_dot}. Applying these shifts to our asymptotic expansions yield equivalent solutions due to the time translation invariance of the EOMs. However, the form of $\Xi=\theta+\mathcal{O}(\lambda)$ is violated unless $\theta_{s}+t_{s}=0$. Therefore, we have one remaining symmetry parameter, which we will denote as $\delta\underline{h}_{1}=\frac{3}{2}t_{s}=-\frac{3}{2}\theta_{s}$ because of the shift it induces in $\underline{h}_1$. In summary, if we apply
\begin{equation}
 \lambda(t)\rightarrow\lambda(t-\frac{2}{3}\delta\underline{h}_{1}) \ ,\qquad\theta(t)\rightarrow\theta(t-\frac{2}{3}\delta\underline{h}_{1})+\frac{2}{3}\delta\underline{h}_{1} \ , \label{eq:h1_transforms}
\end{equation}
then our asymptotic solutions transform as
\begin{align}
 H & \rightarrow\lambda+\lambda^{2}\left(h_{1}(\theta)+\delta\underline{h}_{1}\right)+\lambda^{3}\left(h_{2}(\theta)+2h_{1}(\theta)\delta\underline{h}_{1}+\delta\underline{h}_{1}^{2}\right)+\mathcal{O}(\lambda^{4}) \ , \\
 \Xi & \rightarrow\theta+\lambda\xi_{1}(\theta)+\lambda^{2}\left(\xi_{2}(\theta)+(\xi_{1}(\theta)-\frac{2}{3}\omega_{2})\delta\underline{h}_{1}\right)+\mathcal{O}(\lambda^{3}) \ ,
\end{align}
which is equivalent to a change of $h_{j}$ and $\xi_{j}$. Fixing the value of $\underline{h}_1$ breaks this shift symmetry and therefore acts as an additional RS parameter.

Our results in \sref{sec:Bogoliubov-computation} will be computed with a consistent truncation to render the results explicitly independent of the renormalization scheme. (Detailed proof of this appears in \rref{preplamtheta}.)   This gives us the freedom to choose the RS such that
\begin{equation}
 \int_{0}^{2\pi}h_{\ell}(\theta)d\theta=\int_{0}^{2\pi}\xi_{\ell}(\theta)d\theta=0 \label{eq:elegant_RS}
\end{equation}
for all $\ell\geq1$, which is convenient as the expressions tend to be relatively compact in this scheme. The relevant results to $\mathcal{O}(\lambda^3)$ are given by $\underline{h}_1=0$ and
\begin{align*}
  c_{2} & =\frac{27}{32}-\frac{15\alpha_{3}^{2}}{8}+\frac{3\alpha_{4}}{4} \ ,\quad\omega_{2}=-\frac{27}{16}-\frac{15\alpha_{3}^{2}}{4}+\frac{3\alpha_{4}}{2} \ ,\quad c_{3}=\omega_{3}=0 \ , \\
  H&=\lambda+\frac{3\lambda^{2}}{4}\sin2\theta+\lambda^{3}\left(-3\alpha_{3}\sin\theta+\alpha_{3}\sin3\theta\right)+\lambda^{4}\left(-\frac{21\alpha_{3}}{4}\cos\theta \right. \\ & \hspace*{24pt} \left. + \left(\frac{27}{64}-\frac{3\alpha_{3}^{2}}{16}-\frac{9\alpha_{4}}{8}\right)\sin2\theta-\frac{11\alpha_{3}}{4}\cos3\theta+\left(\frac{81}{256}+\frac{51\alpha_{3}^{2}}{64}+\frac{9\alpha_{4}}{32}\right)\sin4\theta\right) \ , \\
  \frac{a\lambda^{2/3}}{\underline{a}} & =1+\lambda^{2}\left(\frac{9}{32}-\frac{5\alpha_{3}^{2}}{8}+\frac{\alpha_{4}}{4}-\frac{3}{8}\cos2\theta\right)+\lambda^{3}\left(3\alpha_{3}\cos\theta-\frac{9}{16}\sin2\theta-\frac{\alpha_{3}}{3}\cos3\theta\right) \ ,
\end{align*}
where $a$ was found by perturbatively solving $\dot{a} = H a$ after the solution to $H$ was found. Due to the defintion \eref{eq:a_underline_def}, the constant of integration had to be $\underline{a}$ given that $\lambda\to0$ in the far future limit.

We will now derive expressions for $t$ and $\theta$ in terms of $\lambda$ as these are needed for evaluating the $\Omega_{s,k}$ and $n\theta$ terms in \eref{eq:phasePsi}, respectively. Using \erefs{eq:lambda_dot}{eq:theta_dot}, we write
\begin{equation*}
  t=\frac{2}{3}\lambda^{-1}(t)+s-\frac{2}{3}\lambda^{-1}(s)+\int_{\lambda(s)}^{\lambda(t)}dx\left(\frac{1}{\beta_{\lambda}(x)}+\frac{1}{\frac{3}{2}x^{2}}\right)
\end{equation*}
\begin{equation*}
  \theta(t)=\frac{2}{3}\lambda^{-1}(t)+\theta(s)-\frac{2}{3}\lambda^{-1}(s)+\int_{\lambda(s)}^{\lambda(t)}dx\left(\frac{\beta_{\theta}(x)}{\beta_{\lambda}(x)}+\frac{1}{\frac{3}{2}x^{2}}\right)
\end{equation*}
for any $t$ and $s$.
Note that $H^{-1}=\lambda^{-1}-\underline{h}_1-\frac{3}{4}\sin 2\theta + \mathcal{O}(\lambda)$ and $\Xi=\theta+\mathcal{O}(\lambda)$ for any RS, and therefore
\begin{equation*}
  s-\frac{2}{3}\lambda^{-1}(s)=-\frac{2}{3}\left(H^{-1}(s)-\frac{3}{2}s+\frac{3}{4}\sin2\Xi(s)+\underline{h}_{1}\right)+\mathcal{O}(\lambda) \ ,
\end{equation*}
\begin{equation*}
  \theta(s)-\frac{2}{3}\lambda^{-1}(s)=\Xi(s)-s-\frac{2}{3}\left(H^{-1}(s)-\frac{3}{2}s+\frac{3}{4}\sin2\Xi(s)+\underline{h}_{1}\right)+\mathcal{O}(\lambda)\ ,
\end{equation*}
which, along with the definitions of the boundary constants in \erefs{eq:H_underline_def}{eq:Xi_underline_def}, implies that we can take the limit as $s\to+\infty$ to write
\begin{equation}
  \boxed{t(\lambda)=\frac{2}{3}\lambda^{-1}-\frac{2}{3}\left(\underline{H}^{-1}+\underline{h}_{1}\right)+\int_{0}^{\lambda}dx\left(\frac{1}{\beta_{\lambda}(x)}+\frac{1}{\frac{3}{2}x^{2}}\right) \ ,} \label{eq:t_of_lambda}
\end{equation}
\begin{equation}
  \boxed{\theta(\lambda)=\frac{2}{3}\lambda^{-1}+\underline{\Xi}-\frac{2}{3}\left(\underline{H}^{-1}+\underline{h}_{1}\right)+\int_{0}^{\lambda}dx\left(\frac{\beta_{\theta}(x)}{\beta_{\lambda}(x)}+\frac{1}{\frac{3}{2}x^{2}}\right)\ ,} \label{eq:theta_of_lambda}
\end{equation}
given that the $\mathcal{O}(\lambda(s))$ corrections vanish in the limit. The integrals on the right converge and can be expanded as polynomials in $\lambda$. We write
\begin{align}
  t(\lambda) & =\frac{2}{3}\lambda^{-1}-\frac{2}{3}\underline{H}^{-1}+\lambda\left(\frac{9}{16}-\frac{5}{4}\alpha_{3}^{2}+\frac{\alpha_{4}}{2}\right)+\mathcal{O}(\lambda^{3})\ , \\
  \theta(\lambda) & =\frac{2}{3}\lambda^{-1}+\underline{\Xi}-\frac{2}{3}\underline{H}^{-1}+\lambda\left(\frac{27}{16}+\frac{5\alpha_{3}^{2}}{4}-\frac{\alpha_{4}}{2}\right)+\mathcal{O}(\lambda^{3})\ ,
\end{align}
where we used the RS below \eref{eq:elegant_RS}. In this scheme, the $\lambda^2$ terms are exactly zero.

\section{\label{sec:Phase--expansion}Asymptotic expansion of the Bogoliubov complex phase}

In this appendix, we will show that the complex phase of \eref{eq:phasePsi} and its derivatives can be evaluated using only the value of the slowly time-varying $\lambda$. We will detail the calculations of each term to $\mathcal{O}(\lambda^2)$ and then simply state the results to $\mathcal{O}(\lambda^4)$. We have already obtained $\theta(\lambda)$ in \eref{eq:theta_of_lambda}, and therefore we focus remaining items of $\Omega_{s,k}$ and $\widetilde{\mathcal{N}}_k$.

Before expanding in $\lambda$, we make the replacement $k\to\underline{a}\lambda^{-2/3}\sqrt{\mathcal{E}_k^2-m_\chi^2}$, where
\begin{equation}
  \mathcal{E}_k\equiv\sqrt{\frac{k^2}{\underline{a}^2}\lambda^{4/3}+m_\chi^2}
\end{equation}
is treated as $\mathcal{O}(\lambda^0)$, which is justified as $\mathcal{E}_k\sim m_\phi$ at resonance times. This eliminates any fractional powers of $\lambda$ from appearing in our expansion, and is equivalent to expanding in powers of the bookkeeping parameter $\varepsilon$ in \eref{eq:epskappadef} after making the replacements $\lambda\to\varepsilon\lambda$ and $k\to\varepsilon^{-2/3}k$.

To simplify the display of these results, we use \eref{eq:dimensionless_def}, $k \to m_\phi k$, and $m_\chi\to m_\phi m_\chi$ to effectively set $m_\phi=1$. Furthermore, we use the RS defined by \eref{eq:elegant_RS}. We write
\begin{equation}
  R = -3\lambda^{2}(1+3\cos2\theta)-\frac{9}{2}\lambda^{3}\left(\sin2\theta+4\alpha_{3}\left(\cos3\theta-\cos\theta\right)\right)
\end{equation}
\begin{equation}
  \dot{R}=18\lambda^{2}\sin(2\theta)+9\lambda^{3}\left(1+2\cos2\theta+2\alpha_{3}(3\sin3\theta-\sin\theta)\right)
\end{equation}
using $R = - 6\dot{H} - 12H^2$ and the chain rule. Applying this and the results below \eref{eq:elegant_RS} to \eref{eq:tildeN} results in
\begin{align}
\widetilde{\mathcal{N}}_{k}&=\frac{\lambda}{2\mathcal{E}_k^2}+\frac{3\lambda^{2}}{4\mathcal{E}_k^2}\left(1-6\xi+\frac{m_{\chi}^{2}}{2}\right)\sin2\theta+\frac{\lambda^{3}}{2\mathcal{E}_k^2}\left\{ -\frac{3\alpha_{3}}{2}\left(1-6\xi+2m_{\chi}^{2}\right)\sin\theta\right. \nonumber \\ 
&+\frac{3m_{\chi}^{2}}{2\mathcal{E}_{k}^{2}}\left(1-6\xi-\frac{\mathcal{E}_{k}^{2}-m_{\chi}^{2}}{2}\right)\cos2\theta+\alpha_{3}\left(\frac{9(1-6\xi)}{2}+m_{\chi}^{2}\right)\sin3\theta \nonumber \\ 
&\left.+\frac{(1-6\xi)\left(\mathcal{E}_{k}^{2}+2m_{\chi}^{2}\right)+\left(\frac{9}{4}-5\alpha_{3}^{2}+2\alpha_{4}\right)m_{\chi}^{2}\left(\mathcal{E}_{k}^{2}-m_{\chi}^{2}\right)}{4\mathcal{E}_{k}^{2}}\right\}    \ , \label{eq:widetNlambda}
\end{align}
where $m_\chi$ and $\mathcal{E}_k$ are understood as $m_\chi/m_\phi$ and $\mathcal{E}_k/m_\phi$, respectively.

We now explain how to obtain $\Omega_{k}(\lambda,\theta)$. We start by subtracting terms that only contribute a time-independent global phase to $\beta_k$, writing \eref{eq:omega_def} as
\begin{align*}
  \Omega_{k}(t)&=m_{\chi}t+\frac{2m_{\chi}}{3\underline{H}}+\int_{t_{f}}^{t}dt'\left(E_{k}(t')-m_{\chi}\right)+\cancel{\int_{t_{i}}^{t_{f}}dt'E_{k}(t')-m_{\chi}t_{f}-\frac{2m_{\chi}}{3\underline{H}}}\\&=m_{\chi}t+\frac{2m_{\chi}}{3\underline{H}}+\int_{0}^{\lambda(t)}d\lambda'\frac{E_{k}(\lambda',\theta(\lambda'))-m_{\chi}}{\beta_{\lambda}(\lambda')}
\end{align*}
where the slashed out terms are the neglected global phase, and the limit $t_f\to\infty$ was taken in the second line. Crucially, this neglected phase is scheme independent, which ensures the same about the remainder. Using the expression for $t(\lambda)$ in \eref{eq:t_of_lambda}, we reduce this to
\begin{equation*}
  \Omega_{k}(t)=\frac{2m_{\chi}}{3}\left(\lambda^{-1}(t)-\underline{h}_{1}\right)+\int_{0}^{\lambda(t)}d\lambda'\left(\frac{E_{k}(\lambda',\theta(\lambda'))}{\beta_{\lambda}(\lambda')}+\frac{m_{\chi}}{\frac{3}{2}\lambda'^{2}}\right) \ ,
\end{equation*}
with the task being to evaluate the integral on the right. Using the decomposition of \eref{eq:nth_decomposition} on $E_k(t)$, we write the slow and fast components of $\Omega_k(t)$ as
\begin{equation}
 \boxed{\Omega_{s,k}(\lambda)=\frac{2m_{\chi}}{3}\left(\lambda^{-1}-\underline{h}_{1}\right)+\int_{0}^{\lambda}d\lambda'\left(\frac{E_{k}^{(0)}(\lambda')}{\beta_{\lambda}(\lambda')}+\frac{m_{\chi}}{\frac{3}{2}\lambda'^{2}}\right) \ ,} \label{eq:Omega_slow_integral}
\end{equation}
\begin{equation}
 \Omega_{f,k}(t)=\sum_{n\neq0}\int_{\infty}^{t}dt'E_{k}^{(n)}(\lambda(t'))e^{in\theta(t')} \ , \label{eq:Omega_fast_integral}
\end{equation}
respectively. The slow component integral converges due to the $m_\chi$ term, and the fast component integral can be solved perturbatively as follows. We write $\Omega_{f,k}$ as a sum over $\Omega^{(n)}_{k}(\lambda)e^{in\theta}$ for $n\neq0$ and take a time derivative of \eref{eq:Omega_fast_integral}, which yields
\begin{equation*}
  \sum_{n\neq0}\left(\beta_{\lambda}(\lambda)\partial_{\lambda}\Omega_{k}^{(n)}(\lambda)+in\beta_{\theta}(\lambda)\Omega_{k}^{(n)}(\lambda)\right)e^{in\theta}=\sum_{n\neq0}E_{k}^{(n)}(\lambda)e^{in\theta} \ ,
\end{equation*}
where the time dependence is now implicit. For each $n\neq0$, this implies
\begin{equation}
 \boxed{\Omega_{k}^{(n)}(\lambda)=\frac{E_{k}^{(n)}(\lambda)-\beta_{\lambda}(\lambda)\partial_{\lambda}\Omega_{k}^{(n)}(\lambda)}{in\beta_{\theta}(\lambda)} \ ,} \label{eq:Omega_fast_recursive}
\end{equation}
which can be solved \emph{recursively} to obtain $\Omega^{(n)}_k$ as an expansion in powers of $\lambda$. This is because $\beta_\lambda\partial_\lambda\Omega_k^{(n)}$ is always suppressed by an extra power of $\lambda$ relative to $\Omega_k^{(n)}$.

To obtain the $E_k^{(n)}$ components, it is convenient to write
\begin{equation}
  E_{k}=\mathcal{E}_{k}\sqrt{1+\left(\frac{\underline{a}^{2}}{a^{2}\lambda^{4/3}}-1\right)\left(1-\frac{m_{\chi}^{2}}{\mathcal{E}_{k}^{2}}\right)+\frac{1-6\xi}{6}\frac{R}{\mathcal{E}_{k}^2}} \ ,
\end{equation}
in which $k$ was written in terms of $\mathcal{E}_k$. When expanding, it is important to \emph{not} expand the implicit $\lambda$ dependence of $\mathcal{E}_k$, and instead treat it as $\mathcal{O}(\lambda^0)$. Using the results of the RS below \eref{eq:elegant_RS}, we write
\begin{align*}
E_{k}&=\mathcal{E}_{k}-\lambda^{2}\left\{ \frac{1-6\xi+(\frac{9}{8}-\frac{5\alpha_{3}^{2}}{2}+\alpha_{4})(\mathcal{E}_{k}^{2}-m_{\chi}^{2})}{4\mathcal{E}_{k}}+\frac{3(1-6\xi-\frac{\mathcal{E}_{k}^{2}-m_{\chi}^{2}}{2})}{4\mathcal{E}_{k}}\cos2\theta\right\} \\
&+\lambda^{3}\left\{ \frac{3\alpha_{3}(1-6\xi-2(\mathcal{E}_{k}^{2}-m_{\chi}^{2}))}{2\mathcal{E}_{k}}\cos\theta-\frac{3(1-6\xi-\frac{3}{2}(\mathcal{E}_{k}^{2}-m_{\chi}^{2}))}{8\mathcal{E}_{k}}\sin2\theta\right.  \\
&\left.-\frac{3\alpha_{3}(1-6\xi-\frac{2}{9}(\mathcal{E}_{k}^{2}-m_{\chi}^{2}))}{2\mathcal{E}_{k}}\cos3\theta\right\}
\end{align*}
which we can now apply to \erefs{eq:Omega_slow_integral}{eq:Omega_fast_recursive} to solve for $\Omega_k(\lambda,\theta)$. The results are  
\begin{align*}
  \Omega_{s,k}&=\frac{2m_{\chi}}{3\lambda}\hypgeo{2}{1}\left(-\tfrac{3}{4},-\tfrac{1}{2};\tfrac{1}{4};1-\tfrac{\mathcal{E}_{k}^{2}}{m_{\chi}^{2}}\right)+\lambda\left\{ \frac{9-20\alpha_{3}^{2}+8\alpha_{4}}{20}\mathcal{E}_{k}+\right. \\
  &\left.\left(\frac{1-6\xi}{6m_{\chi}}+\frac{9-20\alpha_{3}^{2}+8\alpha_{4}}{80}m_{\chi}\right)\hypgeo{2}{1}\left(\tfrac{1}{2},\tfrac{3}{4};\tfrac{7}{4};1-\tfrac{\mathcal{E}_{k}^{2}}{m_{\chi}^{2}}\right)\right\}  \ , \\
  \Omega_{f,k}&=-\frac{3\lambda^{2}(1-6\xi-\frac{\mathcal{E}_{k}^{2}-m_{\chi}^{2}}{2})}{8\mathcal{E}_{k}}\sin2\theta+\lambda^{3}\left\{ \frac{3\alpha_{3}(1-6\xi-2(\mathcal{E}_{k}^{2}-m_{\chi}^{2}))}{2\mathcal{E}_{k}}\sin\theta+\right.\\&\left.\frac{9(1-6\xi)(\mathcal{E}_{k}^{2}+\frac{m_{\chi}^{2}}{3})-\frac{21\mathcal{E}_{k}^{4}}{2}+9\mathcal{E}_{k}^{2}m_{\chi}^{2}+\frac{3m_{\chi}^{4}}{2}}{16\mathcal{E}_{k}^{3}}\cos2\theta-\frac{\alpha_{3}(1-6\xi-\frac{2}{9}(\mathcal{E}_{k}^{2}-m_{\chi}^{2}))}{2\mathcal{E}_{k}}\sin3\theta\right\} 
\end{align*}
up to $\mathcal{O}(\lambda^2)$ and $\mathcal{O}(\lambda^3)$, respectively. We did not include the $\lambda^3$ term of $\Omega_{s,k}$ as its derivatives are suppressed by extra powers of $\lambda$ relative to the $\lambda^3$ term of $\Omega_{f,k}$. This because only the latter depends on $\theta$, which has an $\mathcal{O}(\lambda^0)$ derivative. In addition, the neglected term includes dependence on $\alpha_5$ and $\alpha_6$, which this appendix does not cover.

\section{\label{sec:obtainingboltzmann}Obtaining the usual collision term}

Consider one of the collision terms of \eref{eq:usual}
\begin{equation*}
C_{2\rightarrow2}(k)=V_{3}^{3}\int\frac{d^{3}k_{2}}{(2\pi)^{3}}\frac{d^{3}p_{1}}{(2\pi)^{3}}\frac{d^{3}p_{2}}{(2\pi)^{3}}\mathcal{S}_{2}(p_{1},p_{2})\left|\langle\chi_{k}\chi_{k_{2}},t_{f}|U(t_{f},t_{i})|\phi_{p_{1}}\phi_{p_{2}},t_{e}\rangle\right|^{2} \ .
\end{equation*}
The matrix element of the box normalized states with box volume $V_{3}$ can be written as
\begin{equation}
\langle\chi_{k}\chi_{k_{2}},t_{f}|U(t_{f},t_{i})|\phi_{p_{1}}\phi_{p_{2}},t_{e}\rangle\approx\frac{(2\pi)^{4}\delta^{(4)}(p_{1}+p_{2}-k_{2}-k)i\mathcal{M}_{2\rightarrow2}}{V_{3}^{4}\sqrt{2E_{p_{1}}}\sqrt{2E_{p_{2}}}{\sqrt{2E_{k}}}\sqrt{2E_{k_{2}}}} \ , \label{eq:matrixelementchange}
\end{equation}
which gives
\begin{align*}
C_{2\rightarrow2}(k)& =\frac{V_{3}^{-1}}{2E_{k}}\int d\Pi(k_{2})d\Pi(p_{1})d\Pi(p_{2})\mathcal{S}_{2}(p_{1},p_{2})\left|(2\pi)^{4}\delta^{(4)}(p_{1}+p_{2}-k_{2}-k)i\mathcal{M}_{2\rightarrow2}\right|^{2}\\
 & =\frac{\Delta t}{2E_{k}}\int d\Pi(k_{2})d\Pi(p_{1})d\Pi(p_{2})\mathcal{S}_{2}(p_{1},p_{2})(2\pi)^{4}\delta^{(4)}(p_{1}+p_{2}-k_{2}-k)\left|\mathcal{M}_{2\rightarrow2}\right|^{2}
\end{align*}
\begin{equation*}
d\Pi(p)\equiv\frac{d^{3}p}{(2\pi)^{3}2E_{p}}
\end{equation*}
where we used Fermi's golden rule and $\Delta t$ is the long-time period defining the asymptotic state region. Integrating over $k$:
\begin{equation}
  \int\frac{d^{3}k}{(2\pi)^{3}}C_{2\rightarrow2}(k)=\Delta t\int\frac{d^{3}p_{1}}{(2\pi)^{3}}\frac{d^{3}p_{2}}{(2\pi)^{3}}\mathcal{S}_{2}(p_{1},p_{2})\left|\frac{\vec{p}_{1}}{E_{\vec{p}_{1}}}-\frac{\vec{p}_{2}}{E_{\vec{p}_{2}}}\right|\int d\sigma(p_{1}p_{2}\rightarrow kk_{2}) \ ,
\end{equation}
\begin{equation}
  \int d\sigma(p_{1}p_{2}\rightarrow kk_{2})=\frac{\int d\Pi(k)d\Pi(k_{2})(2\pi)^{4}\delta^{(4)}(p_{1}+p_{2}-k_{2}-k)\left|\mathcal{M}_{2\rightarrow2}\right|^{2}}{4\left|\vec{p}_{1}E_{\vec{p}_{2}}-\vec{p}_{2}E_{\vec{p}_{1}}\right|} \ ,
\end{equation}
where $d\sigma$ is the differential cross section. Hence, we see that if we take $\mathcal{S}_{2}(p_{1},p_{2})=e^{-E_{1}/T}e^{-E_{2}/T}$, we obtain the usual thermal averaged cross section:
\begin{equation}
\int\frac{d^{3}p_{1}}{(2\pi)^{3}}\frac{d^{3}p_{2}}{(2\pi)^{3}}e^{-E_{1}/T}e^{-E_{2}/T}\left|\frac{\vec{p}_{1}}{E_{\vec{p}_{1}}}-\frac{\vec{p}_{2}}{E_{\vec{p}_{2}}}\right|\int d\sigma(p_{1}p_{2}\rightarrow kk_{2})=\langle \sigma v\rangle n_{1}^{\mathrm{eq}}n_{2}^{\mathrm{eq}} \ ,
\end{equation}
justifying the interpretation of $\mathcal{S}_{2}(p_{1},p_{2})$ as the generalization of the Bose-Einstein statistical factor in \eref{eq:usual}. Note that \eref{eq:matrixelementchange} is one of the key approximations that are being modified as the actual interaction region is not $[t_{e},t_{f}]$ but $[t_{2}-\frac{\delta t}{2},t_{2}+\frac{\delta t}{2}]$.

\section{\label{sec:xyz_coefficients}Coefficients in the Bogoliubov formulas} 

The relevant coefficients for the results of \sref{subsec:Sample-analytic-amplitudes} are listed here. The $x_j$ coefficients that appear in $\mathcal{A}_k^{(2\to2)}$ are
\begin{align*}
  x_{0} & =-1037-6496\alpha_{3}^{2}+960\alpha_{4} \ , \\
  x_{1} & =4\left(425+608\alpha_{3}^{2}+576\alpha_{4}\right) \ , \\
  x_{2} & =4\left(-177+1016\alpha_{3}^{2}+816\alpha_{4}\right) \ ,
\end{align*}
the $y_j^{(n)}$ coefficients are
\begin{align*}
 y_{0}^{(1)} & =919+1080\alpha_{3}^{2}-432\alpha_{4} \ , \\
 y_{1}^{(1)} & =16\left(509-270\alpha_{3}^{2}+108\alpha_{4}\right) \ ,
\end{align*}
\begin{align*}
  y_{0}^{(2)} & =961+4320\alpha_{3}^{2}-1728\alpha_{4} \ , \\
 y_{1}^{(2)} & =2\left(7-2160\alpha_{3}^{2}+864\alpha_{4}\right) \ ,
\end{align*}
\begin{align*}
  y_{0}^{(3)} & =81\left(871+9720\alpha_{3}^{2}-3888\alpha_{4}\right) \ , \\
  y_{1}^{(3)} & =144\left(-521+2430\alpha_{3}^{2}-972\alpha_{4}\right) \ ,
\end{align*}
\begin{align*}
  y_{0}^{(4)} & =-135258+704696\alpha_{3}^{2}+471216\alpha_{4}+2350080\alpha_{3}^{4}-110592\alpha_{3}^{2}\alpha_{4}-331776\alpha_{4}^{2} \ , \\
  y_{1}^{(4)} & =38073+795404\alpha_{3}^{2}-211512\alpha_{4}-587520\alpha_{3}^{4}+27648\alpha_{3}^{2}\alpha_{4}+82944\alpha_{4}^{2}, \\
  y_{2}^{(4)} & =2\left(4323-52180\alpha_{3}^{2}+20616\alpha_{4}\right),
\end{align*}
and the $z^{(n)}$ are given by
\begin{align*}
  z^{(n)} & = \frac{-9+20\alpha_{3}^{2}-8\alpha_{4}}{40}\frac{m_{\chi}}{m_{\phi}}\hypgeo{2}{1}\left(\frac{1}{2},\frac{3}{4};\frac{7}{4};1-\frac{n^2 m_{\phi}^{2}}{4m_{\chi}^{2}}\right)
\end{align*}
for all $n\geq1$, where $\hypgeo{2}{1}$ is the hypergeometric function.

\bibliographystyle{JHEP}
\bibliography{interference}

\end{document}